\documentclass[iop, numberedappendix]{emulateapj}
\usepackage[colorlinks,urlcolor=blue,citecolor=blue,linkcolor=blue]{hyperref} 
\slugcomment{{\sc Accepted for publication in the Astrophysical Journal;} to appear January 2013}
\usepackage{array,placeins}

\newcolumntype{x}[1]{>{\raggedleft\hspace{0pt}}p{#1}}
\newcolumntype{y}[1]{>{\centering\hspace{0pt}}p{#1}}
\newlength{\Oldarrayrulewidth}
\newcommand{\tn}{\tabularnewline}
\newcommand{\Cline}[2]{%
  \noalign{\global\setlength{\Oldarrayrulewidth}{\arrayrulewidth}}%
  \noalign{\global\setlength{\arrayrulewidth}{#1}}\cline{#2}%
  \noalign{\global\setlength{\arrayrulewidth}{\Oldarrayrulewidth}}}

\newcommand{\jone}{($J=1\rightarrow0$) }
\newcommand{\jtwo}{($J=2\rightarrow1$) }
\newcommand{\jthree}{($J=3\rightarrow2$) }
\newcommand{\jnine}{($J=9\rightarrow8$) }

\newcommand{\kms}{km~s$^{-1}$}
\newcommand{\kmssp}{km~s$^{-1}$~}

\newcommand{\Kkmspc}{K~km~s$^{-1}$~pc$^2$}
\def\h2{H$_2$}

\def\c2{cm$^{-2}$}

\shorttitle{EGNoG: Gas Excitation}
\shortauthors{Bauermeister et al.}
\begin{document}

\title{The EGNoG Survey: Gas Excitation in Normal Galaxies at $z\approx0.3$}
\author{A. Bauermeister$^1$, L. Blitz$^1$, A. Bolatto$^2$, M. Bureau$^3$, \\
P. Teuben$^2$, T. Wong$^4$ and M. Wright$^1$}
\affil{$^1$Department of Astronomy, University of California at Berkeley, B-20 Hearst Field Annex, Berkeley, CA 94720, USA\\
$^2$Department of Astronomy and Laboratory for Millimeter-wave Astronomy, University of Maryland, College Park, MD 20742, USA\\
$^3$Sub-department of Astrophysics, Department of Physics, University of Oxford, Denys Wilkinson Building, Keble Road, Oxford OX1 3RH, UK\\ 
$^4$Department of Astronomy, University of Illinois, MC-221, 1002 W. Green Street, Urbana, IL 61801, USA}
\email{amberb@astro.berkeley.edu}
%\keywords{Galaxies:ISM --- Galaxies:evolution ---Stars:Formation}

\begin{abstract}
As observations of molecular gas in galaxies are pushed to lower star formation rate galaxies
at higher redshifts, it is becoming increasingly important to understand the conditions of 
the gas in these systems to properly infer their molecular gas content. The rotational transitions
of the carbon monoxide (CO) molecule provide an excellent probe of the gas excitation conditions
in these galaxies.
%, with the ratio of the CO\jthree to CO\jone line ($r_{31}$) of particular use to distinguish
%starburst galaxies from normal galaxies. 
In this paper we present the results from the gas excitation sample of the 
Evolution of molecular Gas in Normal Galaxies (EGNoG) survey at 
the Combined Array for Research in Millimeter-wave Astronomy (CARMA).
%, which uses CO to trace the
%molecular gas content of 31 normal star-forming galaxies from $z=0.05$ to 0.5. 
This subset of the full EGNoG sample consists of four galaxies at $z\approx0.3$ 
with star formation rates of $40-65$ M$_\odot$ yr$^{-1}$ and stellar masses of $\approx 2 \times 10^{11}$ M$_\odot$.
Using the 3 mm and 1 mm bands at CARMA, 
we observe both the CO\jone and CO\jthree transitions in these four galaxies
in order to probe the excitation of the molecular gas.
We report robust detections of both lines in three galaxies (and an upper limit on the fourth),
with an average line ratio, $r_{31} = L_\mathrm{CO(3-2)}' / L_\mathrm{CO(1-0)}'$, 
of $0.46 \pm 0.07$ (with systematic errors $\lesssim 40\%$), which
implies sub-thermal excitation of the CO\jthree line.
We conclude that the excitation of the gas in these massive, highly star-forming galaxies 
is consistent with normal star-forming galaxies such as local spirals, not starbursting systems like 
local ultra-luminous infrared galaxies. 
Since the EGNoG gas excitation sample galaxies are selected from the
main sequence of star-forming galaxies, we suggest that this result is applicable to
studies of main sequence galaxies at intermediate and high redshifts, %\citep[e.g.][]{Tacconi2010,Daddi2010a}, 
supporting the assumptions made in studies that find molecular gas fractions in star forming galaxies at $z\sim1-2$ 
to be an order of magnitude larger than what is observed locally.
\end{abstract}


\section{Introduction}

In the past decade, molecular gas observations have begun probing the
high redshift universe in a systematic way  using increasingly powerful
millimeter instruments.
The picture that is emerging at redshifts $1-2$ is similar in some respects
to what we see in the local universe.  Sub-mm galaxies (SMGs) are observed to be
undergoing extreme starbursts as a result of major interactions or mergers
\citep[e.g.][]{Sheth2004,Tacconi2008,Engel2010}, equivalent to local ultra-luminous infrared galaxies (ULIRGs)
\citep[e.g.][]{Sanders1986,Solomon1997,DownSol1998}. 
Star-forming galaxies at high redshifts akin to local spirals are becoming accessible as well. 
Recent works (\citealp{Tacconi2010,Daddi2010a}; see also work by \citealp{Baker2004} on a 
$z=2.7$ Lyman break galaxy) suggest that $z\sim1-2$ star-forming galaxies
(with star formation rates (SFRs) of $\approx50-200$ M$_\odot$ yr$^{-1}$)
are scaled up versions of local spirals, forming stars in a steady mode (not triggered by interaction), 
despite hosting star formation rates at the level of 
typical local starburst systems like luminous infrared galaxies (LIRGs) and ULIRGS.

%This comparison brings into focus the evolution of normal star-forming galaxies (SFGs) with redshift. }
While galaxies classified as LIRGs or ULIRGs (by their infrared luminosities only)
have typically been associated with starbursting and merging systems by analogy to galaxies in the local 
universe, it is becoming clear that this connection does not hold at high redshifts.
Morphological studies find that while 50\% of local LIRGs show evidence for major mergers \citep{wang2006}, 
that fraction appears to decrease toward high redshifts: \citet{Bell2005} find that more than half of intensely star-forming galaxies
 at $z\approx0.7$ have spiral morphologies and fewer than 30\% show evidence of strong interaction.
Further, the typical SFR of (normal) star-forming galaxies (like local spirals) increases toward higher redshift.
Star-forming galaxies have been observed to obey a tight relation between stellar mass and star formation rate
out to $z\sim2.5$ \citep[e.g.][]{Brinchmann2004, Noeske2007, Elbaz2007, Daddi2007, Pannella2009, Elbaz2011, Karim2011}.
This `main sequence' defines what is normal for star-forming galaxies as a function of redshift, showing that typical
SFRs increase with redshift. Throughout this work, we use SFG to refer to main sequence (normal)
star-forming galaxies. 

The increase in the star formation rate is mirrored in the molecular gas fraction of these systems. 
While studies of local spirals (e.g. FCRAO survey, \citealt{Young1995}; 
BIMA SONG, \citealt{Regan2001}, \citealt{Helfer2003}; HERACLES, \citealt{Leroy2009}; 
also \citealp{Combes1994,Kuno2007}) 
find average molecular gas fractions $f_\mathrm{mgas} = M_\mathrm{mgas}/(M_*+M_\mathrm{mgas}) \sim$5\% 
(where $M_\mathrm{mgas}$ is the molecular gas mass (including He) and $M_*$ is the stellar mass),
observations of high redshift SFGs suggest molecular gas fractions of 20-80\%, an order
of magnitude higher than local spirals. It is clear that in order to understand these SFGs at $z\sim1-2$,
we must investigate the condition of the molecular gas which is forming stars at such an enhanced rate.
Do these systems truly hold massive reservoirs of molecular gas or is enhanced excitation of the
gas misleading the interpretation of the observations?

The rotational transitions of the carbon monoxide (CO) molecule provide
a direct probe of the excitation of the molecular gas in galaxies. 
Local starburst galaxies and ULIRGs \citep[e.g.][]{Bayet2004,Weiss2005a,Papa2007,Greve2009} and high redshift SMGs and quasars
\citep[e.g.][]{Weiss2005b, Riechers2006, Weiss2007, Riechers2011z2quasars, Riechers2011z3SMGs} show signatures of excited
molecular gas: observed CO line spectral energy distributions peak at J$_\mathrm{upper}>5$
with thermalized lines up to J$_\mathrm{upper}>3$.
In contrast, studies of the Milky Way \citep{Fixsen1999} and local SFGs
\citep[e.g.][]{Mauersberger1999,Yao2003,Mao2010} find a wide spread of excitation conditions, 
with an average that implies less-excited gas, where the J$_\mathrm{upper}=3$ line is already subthermal.
This suggests that the CO\jthree line in particular is an indicator of the star formation
character of a galaxy: `starburst' versus `normal'.
%Thus, the ratio of the CO\jthree and CO\jone line luminosities ($r_{31} = L_{CO32} / L_{CO10} \sim T_{b,32}/T_{b,10}$)
%is a useful diagnostic of the condition of the molecular gas in star-forming galaxies.
%The $J=3$ state is 33 K above the ground level (compared to 5.5 K for the $J=1$ state)
%and the critical density is $\sim10^5$ cm$^{-3}$ (compared to $10^{3.5}$ cm$^{-3}$ for $J=1$).
%
%The discrimination between starburst and normal galaxies is particularly of interest
%for the calculation of molecular gas masses from CO line emission.
%The lower CO rotational lines used as tracers of the total
%molecular gas mass are optically thick, so a locally-calibrated conversion factor ($X_\mathrm{CO}$ or $\alpha_\mathrm{CO}$) 
%is used to calculate the molecular gas mass from the CO line luminosity. Local work suggests
%separate conversion factors for normal star-forming galaxies (a Milky Way-like value $\alpha_\mathrm{CO} = 3.2$ 
%M$_\odot$ (\Kkmspc)$^{-1}$; e.g. \citealt{Dame2001}) and starburst systems (a ULIRG-like value 
%$\alpha_\mathrm{CO} = 0.8$ M$_\odot$ (\Kkmspc)$^{-1}$; \citealt{DownSol1998}). 
%(These $\alpha_\mathrm{CO}$ values do not include Helium.)
%This observational trend is supported by increasingly realistic theoretical modeling \citep[e.g.][]{Narayanan2011, OstrShet2011}.
%While the conversion factor is a complex problem, expected to vary as a function of gas excitation, density, 
%metallicity and radiation field \citep{Shetty2011, Leroy2011}, the measurement of rotation line luminosity ratios
%is a useful first step.

More important still, the ratio of higher rotational lines of CO to the CO\jone line is {\it necessary} 
to translate the observed CO line luminosity to a molecular gas mass using $X_\mathrm{CO}$ or $\alpha_\mathrm{CO}$, 
since this conversion factor is calibrated for the CO\jone luminosity.
Therefore, the measurement of $r_{J1} = L_{\mathrm{CO}(J - (J-1))}' / L_{\mathrm{CO}(1-0)}'$ not only informs our interpretation of 
the current intermediate and high redshift CO line studies, but is critical in the
era of ALMA, which provides an order of magnitude increase in sensitivity, 
making the CO lines in high redshift galaxies more accessible.
The CO\jthree line in particular, observed in the 1mm, 2mm and 3mm bands, 
probes the molecular gas in galaxies at z$\sim0.3-3$. This redshift range is of particular
interest since it includes the peak of the star formation rate density of the universe ($z\sim1-2$) 
and therefore the height of galaxy building.

While several studies have measured $r_{31}$ in the local universe 
\citep{Mauersberger1999,Yao2003,Mao2010,Papa2011},
the measurement of line ratios in intermediate and high redshift galaxies has mostly been limited
to SMGs and quasars (see the review by \citealt{SolomonVandenBout2005}).
Our knowledge of CO line ratios in SFGs at $z\sim1-2$ is limited to one study at $z=1.5$ 
(presented in two papers: \citealt{Dannerbauer2009} and \citealt{Aravena2010}) 
which measures $r_{21}$ in three galaxies and $r_{31}$ in
one galaxy. It is clear that more work is needed to better constrain the line ratios in intermediate
and high redshift SFGs to interpret existing and future data and better understand the state
of the molecular gas in these systems.

As part of the Evolution of molecular Gas in Normal Galaxies (EGNoG) survey, 
we observe both the CO\jone and CO\jthree lines in four galaxies at $z\approx0.3$ 
(the gas excitation sample), more than doubling the number of SFGs at $z>0.1$ in
which CO line ratios have been measured. In this paper, we present the $r_{31}$ values
for the EGNoG gas excitation sample and compare to previous work at
low and high redshifts. We discuss the implications of this measurement for the 
excitation of the molecular gas in the observed galaxies as well as for high-redshift SFGs.

The paper is organized as follows: in Section \ref{sec:EGNOG} we give a brief description
of the EGNoG survey as a whole, describe the selection of the gas excitation sample and 
present the redshifts, SFRs and stellar masses of the galaxies in this work; 
in Section \ref{sec:data} we describe the observations and data reduction; in Section \ref{sec:r31} we present
the analysis of $r_{31}$ in these galaxies; in Section \ref{sec:discussion} we discuss the implications of
this work; and we give some concluding remarks in Section \ref{sec:conclusions}.
The data reduction and measurement of fluxes is discussed in detail in Appendix \ref{sec:datreducandflux}
and moment maps of the detected CO emission are presented in Appendix \ref{sec:mommaps}.
Throughout this work, we use a $\Lambda$CDM cosmology with (h, $\Omega_\mathrm{M}$, $\Omega_{\Lambda}$)
 =  (0.7, 0.3, 0.7).

\section{The EGNoG Survey}
\label{sec:EGNOG}

The EGNoG survey is a key project at the 
Combined Array for Research in Millimeter-wave Astronomy (CARMA).\footnote{CARMA is a 3-band, 
23-element millimeter interferometer jointly operated by the
California Institute of Technology, University of California Berkeley, 
University of Chicago, University of Illinois at Urbana-Champaign, and University of Maryland.}
By observing rotational lines of the $^{12}$CO molecule, the EGNoG survey
traces the molecular gas in 31 galaxies from $z=0.05$ to $0.5$, where the significant evolution from 
gas-rich galaxies at $z\sim 1-2$ \citep{Daddi2010a, Tacconi2010} to relatively gas-poor local galaxies 
(e.g. BIMA SONG, \citealt{Helfer2003}; HERACLES, \citealt{Leroy2009}; CARMA STING, \citealt{Rahman2012}) 
remains almost entirely unobserved.
The full EGNoG sample is split into 4 redshift bins:
bin A, 13 sources, $z=0.05-0.1$; bin B, 10 sources, $z=0.16-0.20$; bin C, 4 sources, $z=0.28-0.32$; 
and bin D, 4 sources, $z=0.47-0.53$. 
The full survey is presented in the forthcoming paper Bauermeister et al. 2013b, {\it in preparation}.

In this paper, 
we present the gas excitation sample (bin C) of the EGNoG survey, the
subset of the full survey sample for which we observe both the CO\jone and CO\jthree lines.
Observations were made using the CARMA 15-element array with 
the 3mm band (single-polarization) for the CO\jone observations 
and the 1mm band (dual-polarization) for the CO\jthree observations.

\begin{table*}[!ht]
\renewcommand{\arraystretch}{1.4}
\centering
\begin{tabular}{|c|c|c|c|c|c|c|}
\hline
EGNoG & SDSS identification & RA & Dec & $z$ & $\log(M_*/M_\odot)$ & SFR  \\
Name & & & & & & (M$_\odot$ yr$^{-1}$) \\
\hline
 C1$^a$ & SDSS J092831.94+252313.9 & 09:28:31.941 & +25:23:13.925 & $0.283020 \pm 0.000022$ & $11.24^{+0.10}_{-0.11}$ & $38.7^{+85.9}_{-25.6}$ \\
 C2 & SDSS J090636.69+162807.1 & 09:06:36.694 & +16:28:07.136 & $0.300622 \pm 0.000010$ & $11.20^{+0.29}_{-0.14}$ & $57.5^{+90.1}_{-21.9}$ \\
 C3 & SDSS J132047.13+160643.7 & 13:20:47.139 & +16:06:43.720 & $0.312361 \pm 0.000014$ & $11.46^{+0.25}_{-0.12}$ & $64.9^{+142.9}_{-28.0}$ \\
 C4 & SDSS J133849.18+403331.7 & 13:38:49.189 & +40:33:31.748 & $0.285380 \pm 0.000015$ & $11.26^{+0.18}_{-0.12}$ & $50.5^{+48.1}_{-15.4}$ \\
\hline
\end{tabular}
\caption{Basic information. Derived quantities ($z$, $M_*$, SFR) are from the MPA-JHU group (see text).\\
$^a$ indicates duplicate source in SDSS: the average value is reported for $z$, $M_*$ and SFR.}
\label{tab:sdss}
\end{table*}

\subsection{Sample Selection}
The galaxies of EGNoG bin C are drawn from the main spectroscopic sample of the 
Sloan Digital Sky Survey (SDSS), Data Release 7 \citep{York2000, Strauss2002, Abazajian2009}.
The galaxies were selected from the parent sample
to be as representative as possible of the main sequence of star-forming galaxies.
The main sequence is the tight correlation between $M_*$ and SFR that has been observed over a large range of redshifts:
e.g. $z\sim0$ \citep{Brinchmann2004}, $z\sim0.2-1$ \citep{Noeske2007}, $z\sim1-2$ \citep{Elbaz2007, Daddi2007, Pannella2009}, 
(see also the summary of recent results in \cite{Dutton2010}).
Building on these observations, a few authors have attempted to describe the main sequence relation 
at $0 \leq z \lesssim 2.5$ with one equation \citep{Bouche2010,Karim2011,Elbaz2011}. 
In order to classify the EGNoG galaxies, we adopt a relation which roughly agrees with the 
relations from \cite{Bouche2010}, \cite{Karim2011} and \cite{Elbaz2011} 
(see Bauermeister et al. 2013b, {\it in preparation}, for a complete description):  
\begin{equation}
sSFR_\mathrm{MS}(\mathrm{Gyr}^{-1}) = 0.07 (1+z)^{3.2} \left(\frac{M_*}{10^{11}\ \mathrm{M}_\odot}\right)^{-0.2} 
\label{equ:myssfr}
\end{equation}
where sSFR$_\mathrm{MS}$ is the specific star formation rate (sSFR$ = \mathrm{SFR} / M_*$) of the 
main sequence (MS) of star-forming galaxies. 
We differentiate `starburst' (SB) from `normal' galaxies according the criteria of \cite{Rodighiero2011}: 
$sSFR_\mathrm{SB} > 4 \times sSFR_\mathrm{MS}$. Therefore, SFGs (normal star-forming galaxies) lie roughly
within a factor of 4 of the main sequence sSFR (which is a function of $M_*$ and $z$) and
starbursts lie at sSFR values larger than 4 times the main sequence sSFR.

In the selection of the EGNoG sample, we apply the following criteria in each redshift bin
in order to identify non-interacting, star-forming galaxies lying as close to the main sequence as possible.
Star-forming galaxies are selected using the  BPT diagram \citep{BPT1981}
criteria from \cite{Kauffmann2003} (rejecting sources with active galactic nuclei).
Obviously interacting galaxies were excluded via visual inspection of the SDSS optical images.
However, some interacting galaxies may be in the sample due to the difficulty of identification at the modest resolution of the SDSS images.
Practical considerations imposed the following further constraints. 
We required a spectroscopic redshift so that the error in the redshift is
small enough to ensure that CO emission would be captured within the observed bandwidth.
We excluded galaxies with SFRs below a minimum value estimated from the instrument 
sensitivity.

\begin{figure}[b]
\centering
\includegraphics[width=0.75\linewidth]{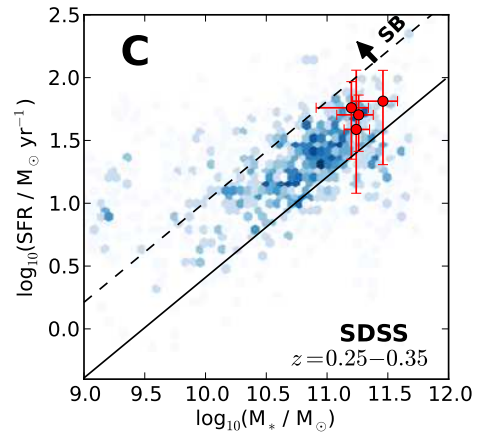}
\caption{Stellar mass versus SFR in EGNoG bin C. 
Red points and error bars show EGNoG galaxies and blue shading shows the logarithm of the 
density (in the $M_*$-SFR plane) of star-forming galaxies from 
the parent dataset at $z=0.25-0.35$
(slightly larger than the EGNoG redshift range so that more 
points may be included to better capture the behavior of the main sequence).
The solid black line indicates the main sequence (Equation \ref{equ:myssfr})
at the average redshift of the bin. The starburst (SB) criterion is indicated by the black dashed line.}
\label{fig:massvsfr}
\end{figure}

The result of the sample selection is illustrated in Figure \ref{fig:massvsfr}, which 
shows stellar mass versus SFR of the EGNoG bin C galaxies (red points). The blue shading
indicates the (logarithm of the) density of points in the $M_*$ - SFR plane for all star-forming galaxies 
(spectroscopic targets only; $\approx 1300$ galaxies) at $z=0.25-0.35$.
We plot galaxies in a slightly larger redshift range than the bin C specification 
in order to better capture the behavior of the main sequence. 
The main sequence of star-forming galaxies at $z=0.3$
is indicated by the solid black line, with the starburst cutoff indicated by the dashed black line.
In this plot, the low-mass, low-SFR end of the main sequence is sparsely sampled 
due to the limited number of spectroscopically targeted, low-SFR, star-forming galaxies
in the SDSS at $z\approx0.3$. 
While the EGNoG bin C galaxies are high-$M_*$, high-SFR galaxes, they lie
within the expected scatter of the main sequence (as defined in Equation \ref{equ:myssfr}), 
and are not classified as starburst galaxies according to the prescription
from \cite{Rodighiero2011} (presented above).

\subsection{Redshifts, Stellar Masses and Star Formation Rates}
\begin{figure}[t]
\centering
\begin{minipage}[h]{0.45\linewidth}
\centering
\includegraphics[width=\linewidth]{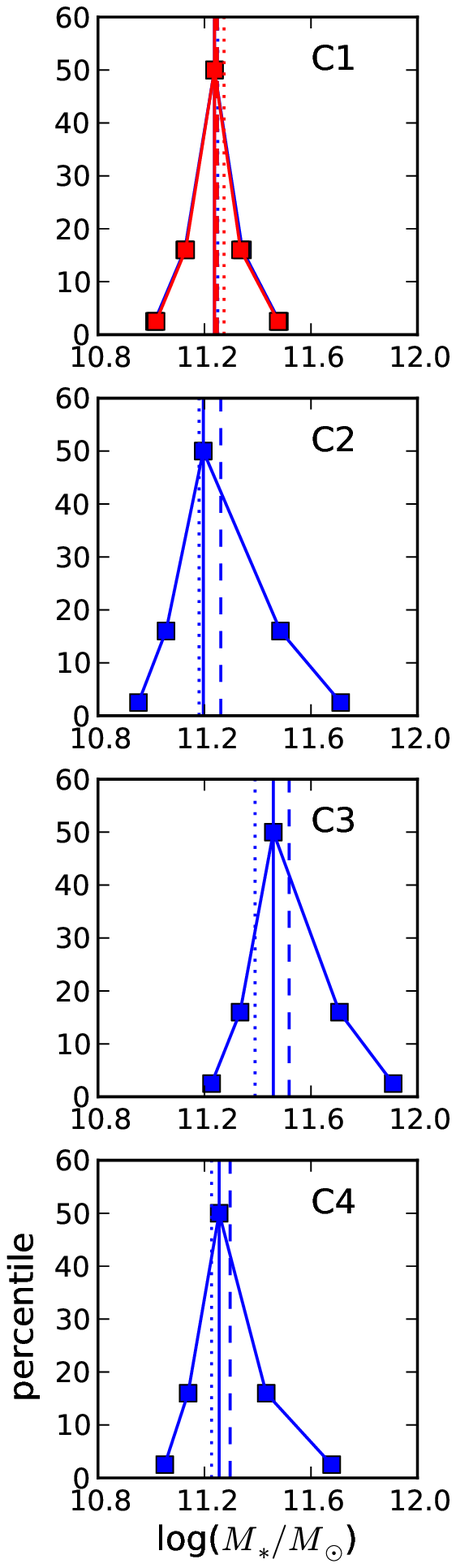}
\end{minipage}
\begin{minipage}{0.45\linewidth}
\centering
\includegraphics[width=\linewidth]{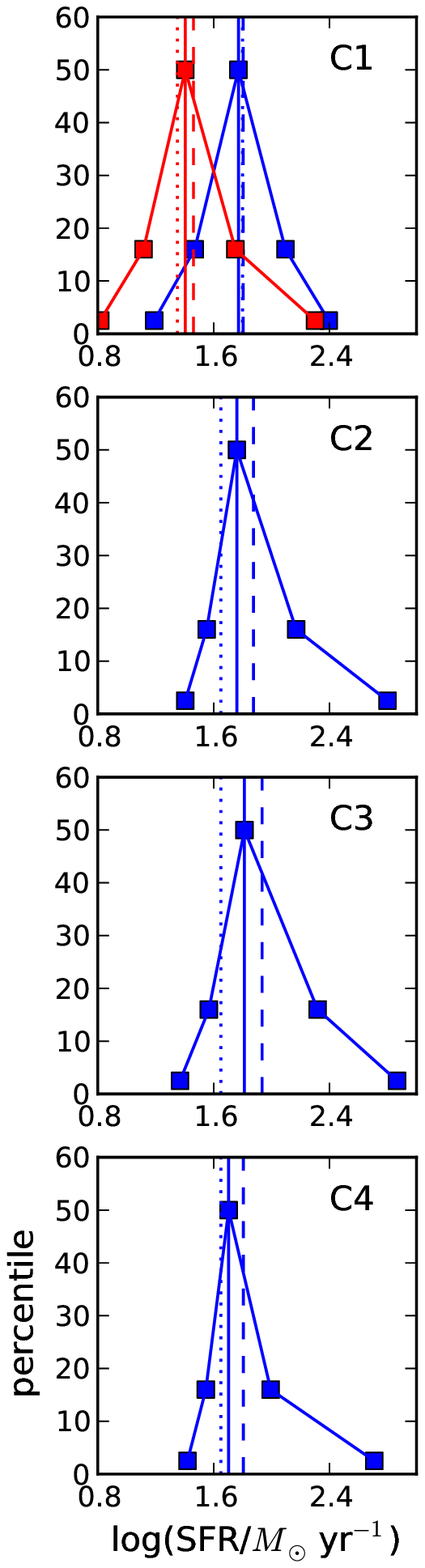}
\end{minipage}
\caption{Probability distribution functions for the stellar masses (left panel) and star formation
rates (right panel)  from the MPA-JHU group for bin C sources.
Data points give the 2.5, 16, 50, 84 and 97.5 percentiles of the distributions. The median, 
mean and mode are indicated by the vertical solid, dashed and dotted lines respectively. 
In cases where a duplicate exists (source C1), the duplicate values are plotted in red. }
\label{masssfrPDFs}
\end{figure}

Spectroscopic redshifts are from David Schlegel's {\tt spZbest} files produced by 
the Princeton-1D code, {\tt specBS}\footnote{See \url{http://spectro.princeton.edu/} for more information}.
The stellar masses and SFRs of galaxies in 
the SDSS DR7 are provided by the Max-Planck-Institute for Astrophysics - John 
Hopkins University group (\url{http://www.mpa-garching.mpg.de/SDSS}). 
Stellar masses are derived by fitting SDSS ugriz photometry to a grid of models
spanning a wide range of star formation histories. This method is found
to compare quite well with the \cite{Kauffmann2003} methodology
using spectral features (more detail on this comparison is found on the website above).
Star formation rates are derived by fitting the fluxes of no less than 5 emission lines
using the method described in \cite{Brinchmann2004}.
Both stellar masses and star formation rates are derived using a Bayesian analysis,
producing probability distributions of each quantity for each galaxy. These distributions
for the four bin C galaxies are shown in Figure \ref{masssfrPDFs}. We take the median of 
the distribution, with errors indicated by the 16th and 84th percentile points. In the case
where a duplicate SDSS source exists (as a result of SDSS automated source-finding), 
we take the average of the two median values and use
the lowest(highest) 16th(84th) percentile value to indicate the negative(positive) error. 
The redshifts, stellar masses and star formation rates (with errors) are given in Table \ref{tab:sdss}.

\begin{figure*}[t]
\centering
\begin{minipage}[h]{1.93in}
\includegraphics[height=3.12in]{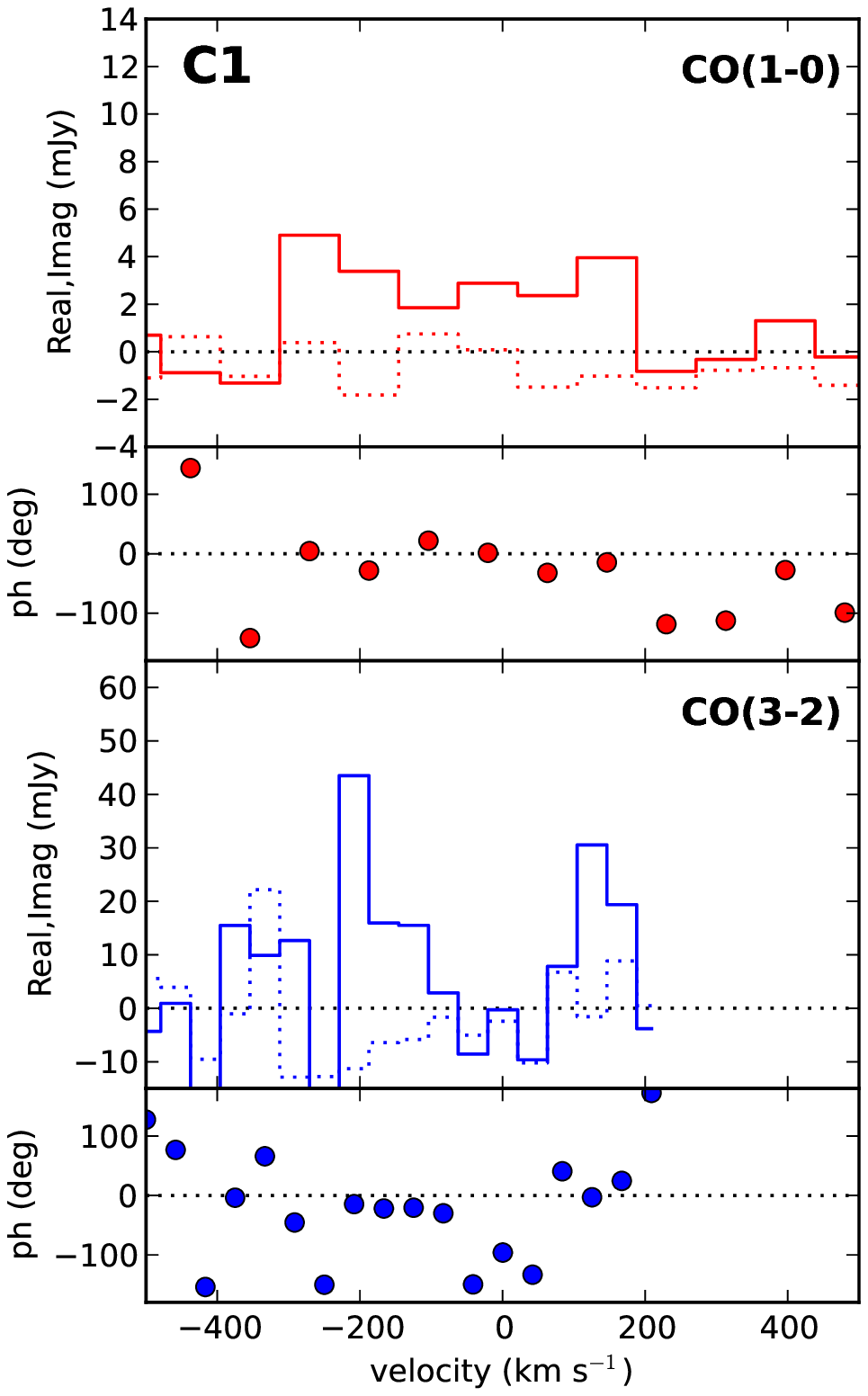}
\end{minipage}
\begin{minipage}[h]{1.58in}
\includegraphics[height=3.12in]{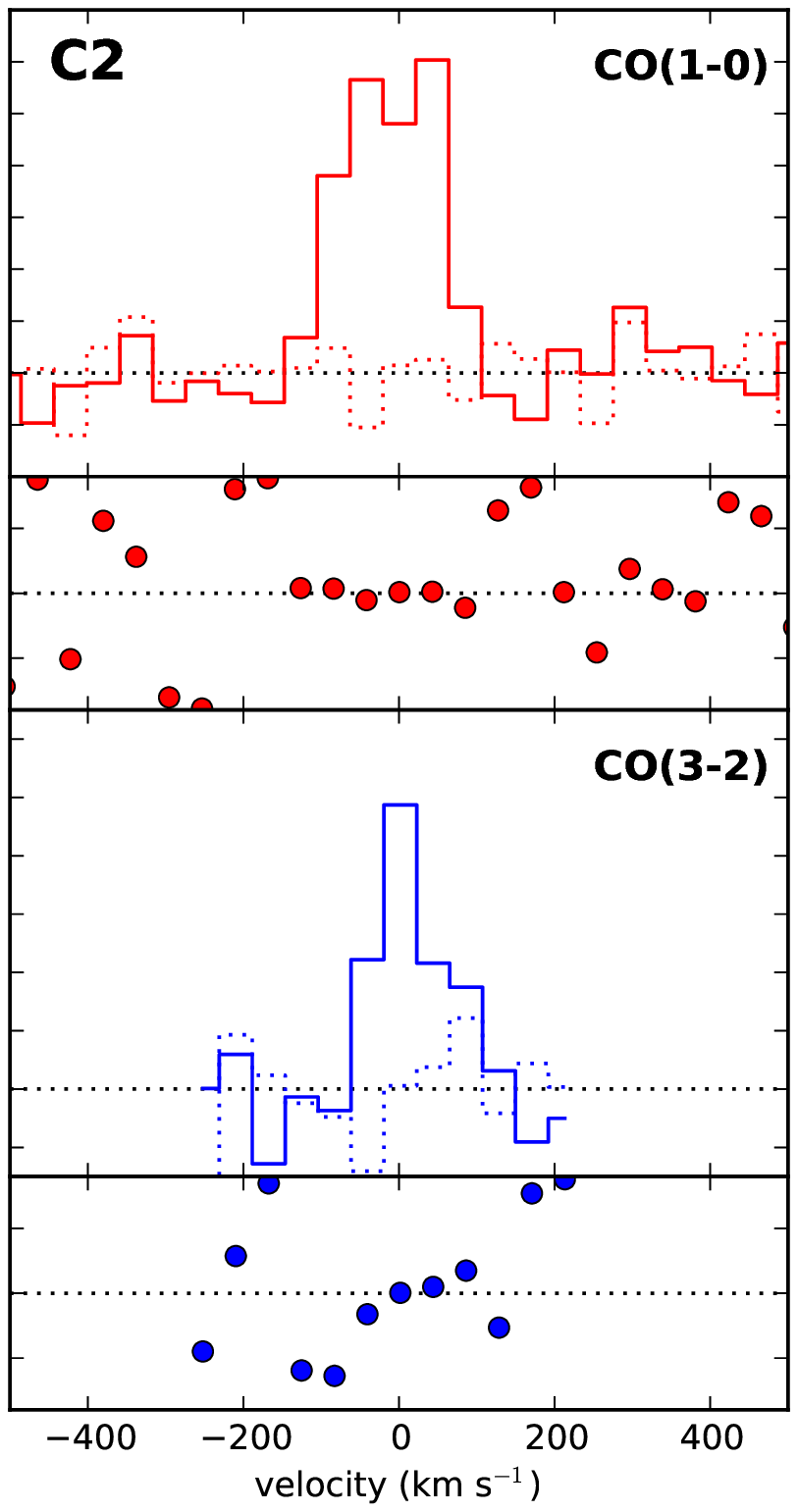}
\end{minipage}
\begin{minipage}[h]{1.58in}
\includegraphics[height=3.12in]{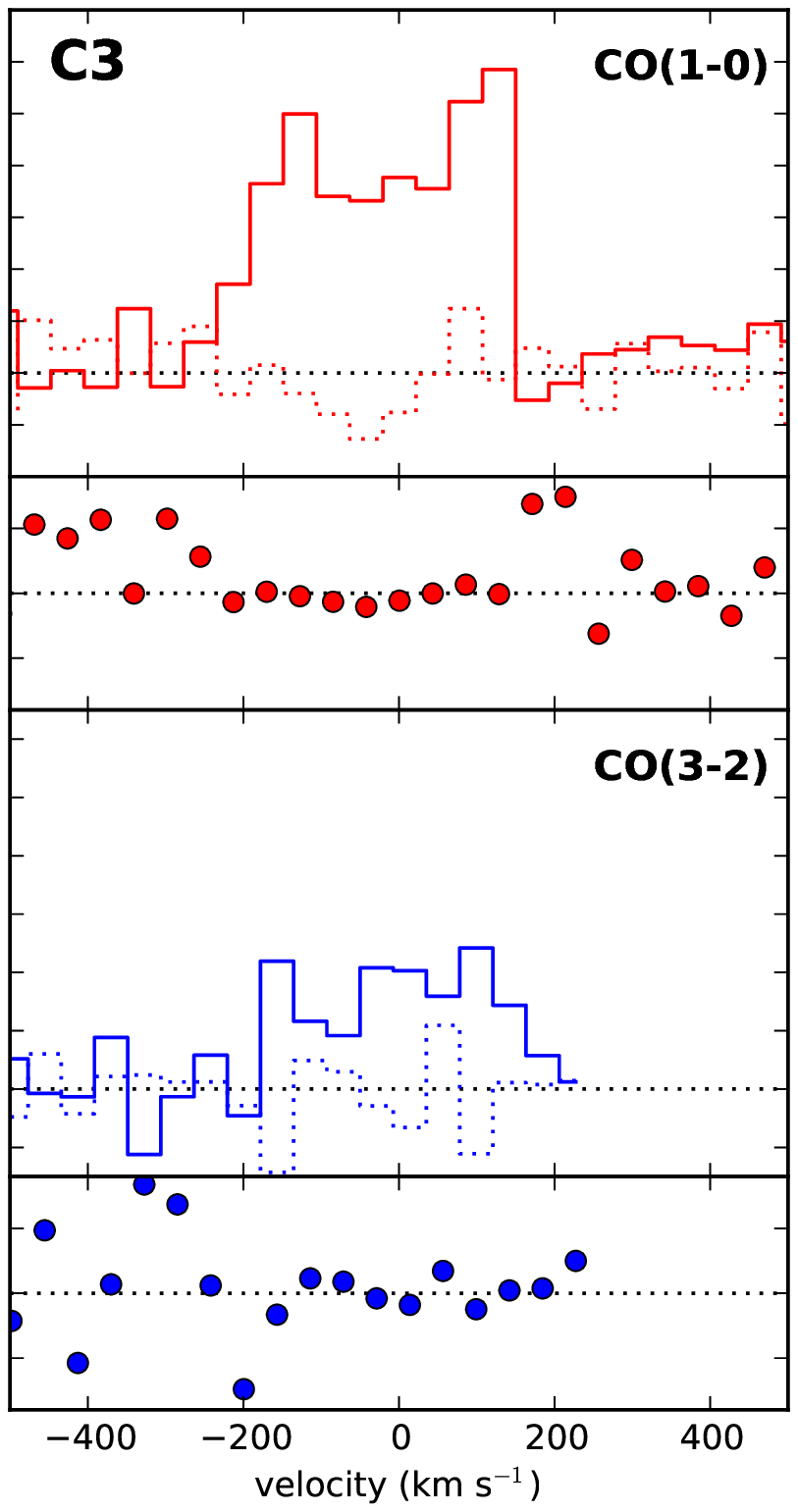}
\end{minipage}
\begin{minipage}[h]{1.58in}
\includegraphics[height=3.12in]{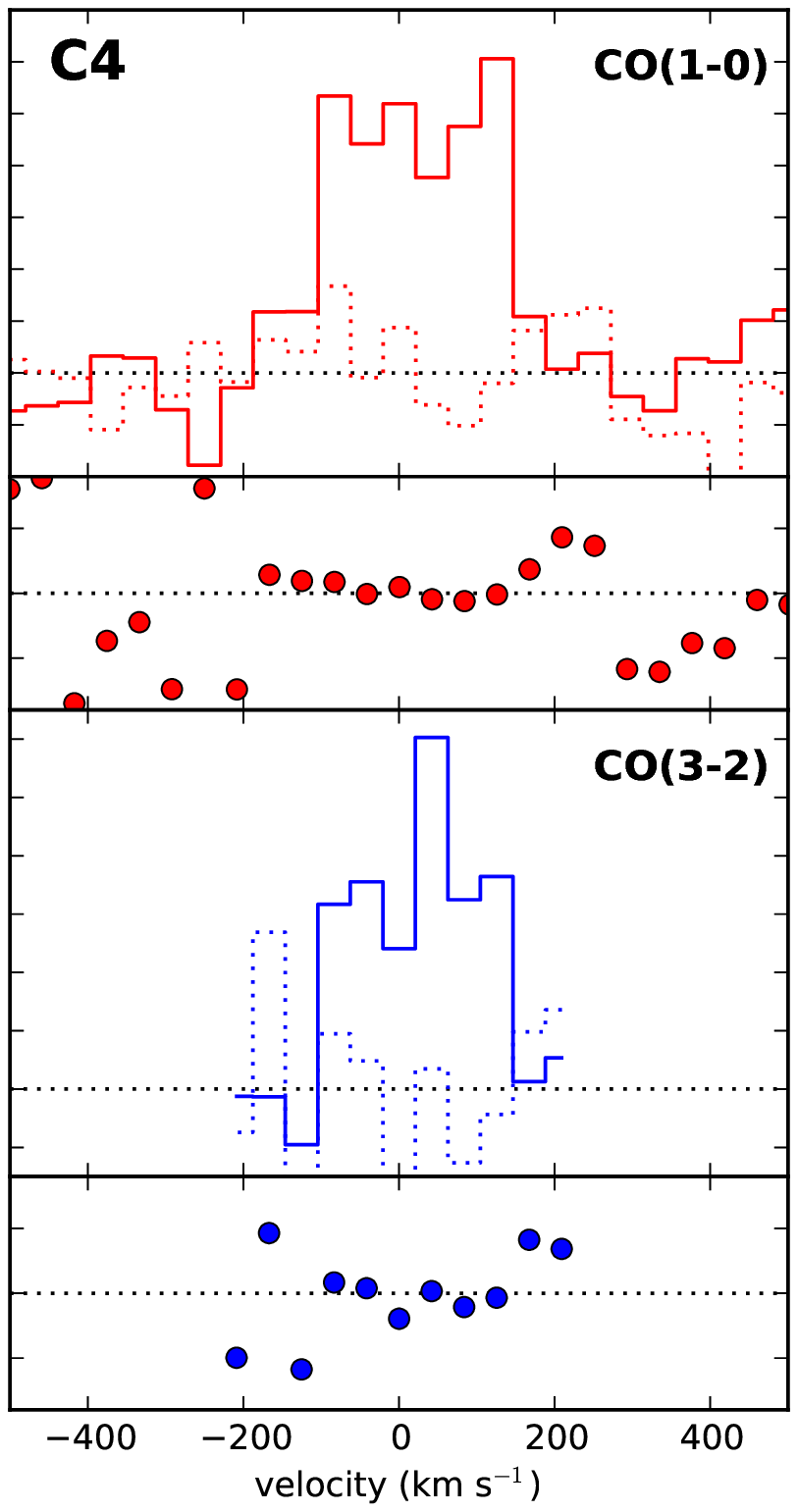}
\end{minipage}
\caption{Vector-averaged {\it Real} (solid) and {\it Imaginary} (dotted) amplitudes (Real,Imag in mJy beam$^{-1}$) 
and phase (ph, degrees) of the calibrated $uv$ data versus velocity (\kms)
for the CO\jone (red, top panel) and CO\jthree (blue, bottom panel) transitions 
in each of the four sources (indicated in upper left corner). With the higher resolution of the CO\jthree
data, we observe the peak of the CO\jthree emission to be offset ($<1.5\arcsec \sim 6$ kpc)
from the centers of galaxies C1 and C2. 
In these cases, we calculate the $uv$ spectra at the peak of the CO\jthree emission.
For the CO\jone line of galaxy C1, we average 2 channels together
in order to increase the signal to noise ratio. 
Large {\it Real} amplitudes (the {\it Imaginary} part shows the noise) coincident with 
phases of $\approx 0$ over multiple velocity channels indicate a detection.}
\label{fig:uvspec}
\end{figure*}

\section{CARMA Observations}
\label{sec:data}

The four bin C galaxies were observed in two rotational transitions of the CO molecule: 
CO\jone $\nu_\mathrm{rest} = 115.3$ GHz, $\nu_\mathrm{obs} \sim 88$ GHz
and CO\jthree $\nu_\mathrm{rest} = 345.9$ GHz, $\nu_\mathrm{obs} \sim 266$ GHz. 
At each frequency, each galaxy was observed over several different days.
Each dataset includes observations of a nearby quasar for phase calibration (taken every 15-20 minutes), 
a bright quasar for passband calibration and either Mars or MWC349 for flux calibration (in most cases). 

The reduction of all observations for this survey was carried out within the 
EGN\footnote{\url{http://carma.astro.umd.edu/wiki/index.php/EGN}} data reduction infrastructure
(based on the MIS pipeline; \citealp{PoundTeuben2012}) using the Multichannel Image Reconstruction, 
Image Analysis and Display (MIRIAD; \citealp{MIRIAD}) package for radio interferometer data reduction.
Our data analysis also used the {\tt miriad-python} software package \citep{miriadpython}.
The data were flagged, passband-calibrated and phase calibrated in the standard way. Final images
were created using {\tt invert} with {\tt options=mosaic} in order to properly handle and correct for the three different
primary beam patterns. All observations are single-pointing. We describe the CO\jone and CO\jthree observations
individually below. A full description of the data reduction and flux measurement is given in Appendix \ref{sec:datreducandflux}.
%The details of the individual observations (date, length, weather, calibrators, etc.) are given in Tables
%\ref{tab:3mmobs} and \ref{tab:1mmobs}.

\subsection{CO\jone}
The CO\jone transition for the galaxies presented here lies in the 3 mm band of CARMA
(single-polarization, linearly polarized feeds).
The wide bandwidth of the CARMA correlator allowed us to observe both the $^{12}$CO\jone
and the $^{13}$CO\jone lines simultaneously.
The $^{12}$CO ($^{13}$CO) line was observed with
five (three) overlapping 500 MHz bands, covering $\approx$ 7000 (5000) \kmssp total, 
at 42 \kmssp resolution, with 3-bit sampling. 
These observations were carried out from August to November 2011 in CARMA's D configuration, with 
$11-150$ m baselines yielding a typical synthesized beam of 
$4.8\arcsec \times 3.9\arcsec$ at 88 GHz.
Each galaxy was observed for 20 to 30 hours (time on-source) in moderate to good weather
conditions for 3mm observation, yielding final images with rms noise of $\approx 1.2$ mJy beam$^{-1}$
in a 42 \kmssp channel. The flux scale in each dataset is set by the flux of the phase
calibrator, which is determined from the flux calibrator. For these data, we use phase calibrators
0854+201 (4 Jy, 9\% linearly polarized), 1357+193 (0.8 Jy) and 1310+323 (1.7 Jy). 

All four galaxies were clearly detected in the CO\jone line, as shown in Figure \ref{fig:uvspec}. The
top panels show the vector-averaged {\it Real} and {\it Imaginary} 
amplitudes and phase of the calibrated $uv$ data
versus velocity for the CO\jone line. For a compact source at the center of the field of view, the
{\it Real} part shows the signal without a noise bias, and the {\it Imaginary} part shows the noise.
In all four cases, we see coherent emission (larger {\it Real amplitudes} coincident with noise-like {\it Imaginary}
amplitudes and phases of $\approx 0$) over multiple velocity channels, indicative of a detection. 
We report no detection of the $^{13}$CO\jone line, which is expected
to be weaker than the $^{12}$CO\jone line by a factor of 7 to 17 \citep{RickardBlitz1985}.

\subsection{CO\jthree}
The CO\jthree transition at the redshift of the galaxies discussed here lies in the 
CARMA 1 mm band (dual-polarization, circularly polarized feeds).
We again observed both the $^{12}$CO\jthree and the $^{13}$CO\jthree lines simultaneously.
The $^{12}$CO ($^{13}$CO) line was observed with
three (one) overlapping 500 MHz bands, covering $\approx$ 1500 (550) \kmssp total, 
at 14 \kmssp resolution, with 3-bit sampling. 
These observations were carried out in CARMA's E and D configuration during August 2011
and April 2012, respectively. 
Source C4 was observed entirely in E configuration, while the other three sources
were observed mostly in D configuration. The E configuration has 
$8-66$ m baselines yielding a typical synthesized beam of 
$3.2\arcsec \times 2.5\arcsec$ at 266 GHz.
The D configuration has $11-150$ m baselines yielding a typical synthesized beam of 
$1.7\arcsec \times 1.5\arcsec$ at 266 GHz.
Each galaxy was observed for 2 to 6.5 hours (time on-source) in good weather
conditions for 1mm observation, yielding final images with rms noise of 5-10 mJy beam$^{-1}$
in a 42 \kmssp channel. For these data, we use phase calibrators
0854+201 (2 Jy in August 2011, 4 Jy in April 2012), 1224+213 (0.6 Jy) and 1310+323 (0.6 Jy). 

Sources C2, C3 and C4 were detected at the $\approx5\sigma$ level in the CO\jthree line.
However, C1, with its wide velocity profile (observed in the CO\jone line), was only marginally detected and we
give only an upper limit on the CO\jthree flux. The vector-averaged {\it Real} and {\it Imaginary} amplitudes 
and phase of the calibrated $uv$ data versus velocity for the CO\jthree line
are shown in the bottom panels of Figure \ref{fig:uvspec}.
The peak of the CO\jthree emission in galaxies
C1 and C2 is offset from the nominal center of each galaxy, so the $uv$ spectra 
are calculated at this slightly offset ($<1.5\arcsec \sim 6$ kpc)) position.
We report no detection in the $^{13}$CO\jthree line for any galaxy.

\subsubsection{Source C1 Upper Limit}
Due to the wide integrated velocity profile of source C1, the CO\jthree line was only marginally detected. The channel maps
did not show evidence of a source upon visual inspection, but an integrated spectrum made of a circular region $4.5\arcsec$
in radius at the center of the image suggests a $3\sigma$ detection. The line flux and other quantities are 
calculated from this spectrum, over the velocities
of the CO\jone line for this galaxy. These values should be taken as an upper limit on the true flux. 
%This is indicated in Table \ref{tab:COprop}.
The error in the flux measurement ($S_\mathrm{CO}$, Jy \kms) is calculated from the pixel noise, $\sigma_\mathrm{p}$ (Jy beam$^{-1}$): 
$\sigma_{S_\mathrm{CO}} = \sigma_\mathrm{p} \ \delta V \sqrt{N_\mathrm{p} / N_\mathrm{eq}}$
where $\delta V$ is the channel width (\kms), $N_\mathrm{p}$ is the number of pixels summed and $N_\mathrm{eq}$ is the number of
pixels equivalent to the beam. 

\begin{table*}[!t]
\centering
\begin{tabular}{| c | c | x{20pt}@{$\pm$}x{16pt} | x{35pt}@{$\pm$}x{16pt} | x{23pt}@{$\pm$}x{16pt} | x{20pt}@{$\pm$}x{16pt} | x{28pt}@{$\pm$}x{16pt} |}
\hline
Name & Transition & \multicolumn{2}{c|}{$S_\mathrm{CO}$}  & \multicolumn{2}{c|}{$L_\mathrm{CO}'$} & \multicolumn{2}{c|}{$v_\mathrm{center}$} & \multicolumn{2}{c|}{$\Delta V$} & \multicolumn{2}{c|}{Non-matched $r_{31}^b$} \tn
 &   & \multicolumn{2}{c|}{(Jy \kms)} & \multicolumn{2}{c|}{($10^9$ \Kkmspc)} & \multicolumn{2}{c|}{(\kms)} & \multicolumn{2}{c|}{(\kms)} &  \multicolumn{2}{c|}{ } \tn
\hline
C1 & CO\jone & $2.05$&$0.34$ & $8.25$&$2.83$ & $-84.9$&$18.1$ & $542.2$&$41.7$ & $0.49^a$&$0.26$ \tn
 & CO\jthree & $9.02^a$&$2.65$ & $4.03^a$&$1.69$ & $-49.9$&$7.3$ & $542.3$&$23.4$ & \multicolumn{2}{c|}{ } \tn
\hline
C2 & CO\jone & $2.35$&$0.10$ & $10.70$&$3.24$ & $-6.7$&$1.2$ & $253.7$&$42.3$ & $0.40$&$0.18$ \tn
 & CO\jthree & $8.39$&$1.12$ & $4.25$&$1.40$ & $19.5$&$8.2$ & $169.2$&$42.3$ & \multicolumn{2}{c|}{ } \tn
\hline
C3 & CO\jone & $4.39$&$0.09$ & $21.70$&$6.53$ & $-24.7$&$5.7$ & $384.0$&$42.3$ & $0.34$&$0.15$ \tn
 & CO\jthree & $13.46$&$1.56$ & $7.39$&$2.38$ & $-14.0$&$7.8$ & $426.7$&$42.7$ & \multicolumn{2}{c|}{ } \tn
\hline
C4 & CO\jone & $3.01$&$0.11$ & $12.30$&$3.72$ & $17.3$&$13.0$ & $292.5$&$41.8$ & $0.44$&$0.19$ \tn
 & CO\jthree & $11.88$&$1.22$ & $5.41$&$1.72$ & $33.5$&$5.0$ & $250.7$&$41.8$ & \multicolumn{2}{c|}{ } \tn
\hline
\end{tabular}
\caption{Properties of the CO emission (described in Section \ref{sec:derivedprop}). For each galaxy, 
the CO line flux ($S_\mathrm{CO}$), luminosity ($L_\mathrm{CO}'$, see Equation \ref{LCOdef}), 
central velocity ($v_\mathrm{center}$) and full velocity
width ($\Delta V$) are given for the CO\jone and CO\jthree lines. The ratio of the line luminosities ($r_{31}$)
%(Equation \ref{r31equ}) 
for each galaxy is given in the last column.\\
$^a$ CO\jthree values for source C1 are upper limits.\\
$^b$ $r_{31}$ reported here is calculated from the total line luminosities (see Section \ref{sec:totr31}). We perform a
more careful calculation, matching the 1 and 3mm data, in Section \ref{sec:carefulr31}.}
\label{tab:COprop}
\end{table*}

\subsection{Derived Properties of the CO Emission}
\label{sec:derivedprop}
Table \ref{tab:COprop} presents the quantities we derive from the CO emission images.
%The derivation of the line flux ($S_{CO}$ in Jy \kms) is described in Section \ref{sec:fluxest}. 
The CO line luminosity is calculated from the line flux ($S_\mathrm{CO}$ in Jy \kms, calculated 
as described in Section \ref{sec:fluxest}) following 
\begin{equation}
\label{LCOdef}
L_\mathrm{CO}' = 3.25 \times 10^7 \ S_\mathrm{CO} \ \nu_\mathrm{obs}^{-2} \ r_\mathrm{com}^2 (1+z)^{-1}
\end{equation}
\citep[see the review by ][]{SolomonVandenBout2005},
where $\nu_\mathrm{obs}$ is in GHz and $r_\mathrm{com}$ is the comoving distance in Mpc. The units
of $L_\mathrm{CO}'$ are \Kkmspc. We report the measurement error 
for $S_\mathrm{CO}$. The error reported for $L_\mathrm{CO}'$ includes both the measurement error
of $S_\mathrm{CO}$ and a 30\% systematic error, added in quadrature (see Section \ref{sec:fluxest} for more details).

The center velocity, $v_\mathrm{center}$, is the flux-weighted average
velocity of the galaxy-integrated spectrum ($v=0$ at the redshift in Table \ref{tab:sdss}). The error reported
is the standard deviation of the $v_{center}$ values found with the three flux measurement methods and
different channel averaging described in Section \ref{sec:fluxest}. 
The reported velocity width ($\Delta V$) is the full width of the emission, where `source' velocity channels are
selected by eye. We give the velocity width of a single channel as the error.
The line ratio $r_{31}$ is given by 
\begin{equation}
\label{r31equ}
r_{31} = L_\mathrm{CO(3-2)}' / L_\mathrm{CO(1-0)}'
\end{equation}
and the error is calculated by propagating the
errors of the individual line luminosities.

Moment maps (discussed in Section \ref{sec:uvsamp} and displayed in Appendix \ref{sec:mommaps}) are created from
the $2\sigma$ clip smooth mask (see Section \ref{sec:fluxest}). Moment 0 (total
intensity) maps are a simple sum of the masked images in the `source'
velocity channels. Moment 1 (intensity-weighted mean velocity) maps are 
produced by summing the masked image multiplied by the velocity in each channel then
normalizing by the moment 0 value.

\section{Analysis}
\label{sec:r31}

We first present our total $r_{31}$ measurements (using the total flux
observed in each line) in Section \ref{sec:totr31}. All four observed galaxies are discussed. 
However, the total $r_{31}$ values use line luminosities calculated completely independently
of one another, making no attempt at matching the spectral or spatial resolution. 
In Section \ref{sec:uvsamp}, we carefully match the CO\jone and CO\jthree data to each other
as well as possible, in both spectral and spatial resolution.
In Section \ref{sec:carefulr31} we re-calculate $r_{31}$ using both the natural and adjusted spatial
resolution, with matched spectral resolution in both cases.
In Section \ref{sec:fittingr31} we investigate the radial dependence of $r_{31}$ by fitting
two-dimensional Gaussians to the total intensity maps in order to derive peak intensities, which are
independent of spatial resolution.
The analysis in Sections \ref{sec:uvsamp}-\ref{sec:fittingr31} is performed only for the three galaxies that are 
detected significantly in both lines (C1 is excluded due to marginal detection of the CO\jthree line).

\subsection{Total $r_{31}$}
\label{sec:totr31}

As a first step, we calculate $r_{31}$ in the simplest way possible, taking the ratio of the
total line luminosities. Of the four galaxies observed, we detect three in both lines, yielding $r_{31}$ 
values (see Table \ref{tab:COprop}) of 0.34 to 0.44 with an average value of 0.39. 
For source C1 we derive an upper limit on $r_{31}$ of 0.49, which is consistent with the other values. 

These $r_{31}$ values carry the caveat that they are determined from line luminosities that are calculated
independently of one another. The `source' velocity channels for each line are selected by eye for the 
most convincing detection.  Further, the CO\jone and CO\jthree lines
are observed at different frequencies in the same configuration (except source C4).
This yields different sampling of the $uv$-plane, which means that
the two maps are sensitive to slightly different spatial scales.
We re-calculate $r_{31}$ more carefully for C2, C3 and C4 in the following sections.

\subsection{Matching the 1 and 3 mm Data}
\label{sec:uvsamp}

In this section, we attempt to standardize the flux calculated for each line as much as possible
by matching both spatial and spectral resolution.
% as well as choosing appropriate source region sizes (for the calculation of the flux). 
First, we use the same spectral 
resolution in both the CO\jone and CO\jthree data, selecting the starting channel
so that the image velocity channels line up exactly.

\begin{table}[b]
\vspace{0.3cm}
\centering
\begin{tabular}{| r | r | c | r | c | r |}
\hline
Name, & \multicolumn{2}{c|}{All $uv$ data} & \multicolumn{2}{c|}{Matched $uv$ data} & Flux \\
\cline{2-5}
Trans. & $S_\mathrm{CO}$ & Beam & $S_\mathrm{CO}$ & Beam & \% Diff. \\ 
\hline
C2 1-0 &  2.30 & $4.90\arcsec \times 4.08\arcsec$ &  2.15 & $4.54\arcsec \times 3.75\arcsec$ &  -6.5\% \\
   3-2 &  7.06 & $1.70\arcsec \times 1.56\arcsec$ &  7.00 & $3.20\arcsec \times 2.84\arcsec$ &  -0.8\% \\
\hline
C3 1-0 &  4.29 & $4.97\arcsec \times 4.06\arcsec$ &  3.71 & $4.62\arcsec \times 3.77\arcsec$ & -13.5\% \\
   3-2 & 13.75 & $1.71\arcsec \times 1.46\arcsec$ & 11.61 & $3.12\arcsec \times 3.02\arcsec$ & -15.6\% \\
\hline
C4 1-0 &  2.95 & $4.99\arcsec \times 4.02\arcsec$ &  2.76 & $4.62\arcsec \times 3.83\arcsec$ &  -6.4\% \\
   3-2 & 12.79 & $3.33\arcsec \times 2.41\arcsec$ & 12.01 & $3.55\arcsec \times 3.08\arcsec$ &  -6.1\% \\
\hline
\end{tabular}
\caption{Comparison of $uv$ data selections. Measured CO line fluxes ($S_\mathrm{CO}$ in Jy \kms) and beam sizes 
for images made using all $uv$ data and matched $uv$ data. 
The fluxes reported here are calculated using the $2\sigma$
smooth masking technique discussed in Section \ref{sec:fluxest}.
The percent difference in the measured flux is defined as $100 (S_\mathrm{CO,match} - S_\mathrm{CO,all})
/S_\mathrm{CO,all}$. }
\label{tab:uvselfluxes}
\end{table}

\begin{figure*}[t]
\centering
\begin{minipage}[h]{0.32\linewidth}
\includegraphics[width=\linewidth]{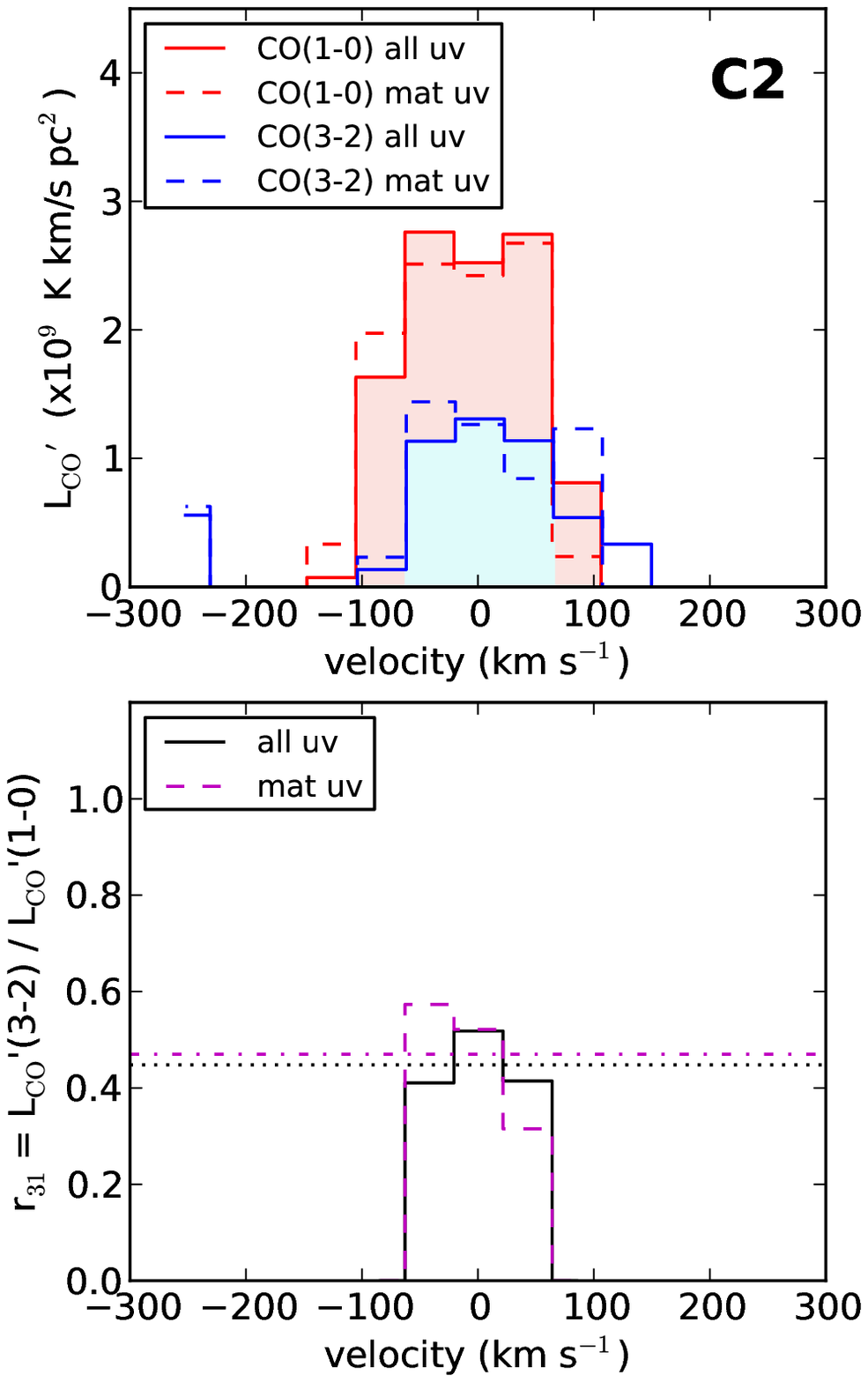}
\end{minipage}
\begin{minipage}[h]{0.32\linewidth}
\includegraphics[width=\linewidth]{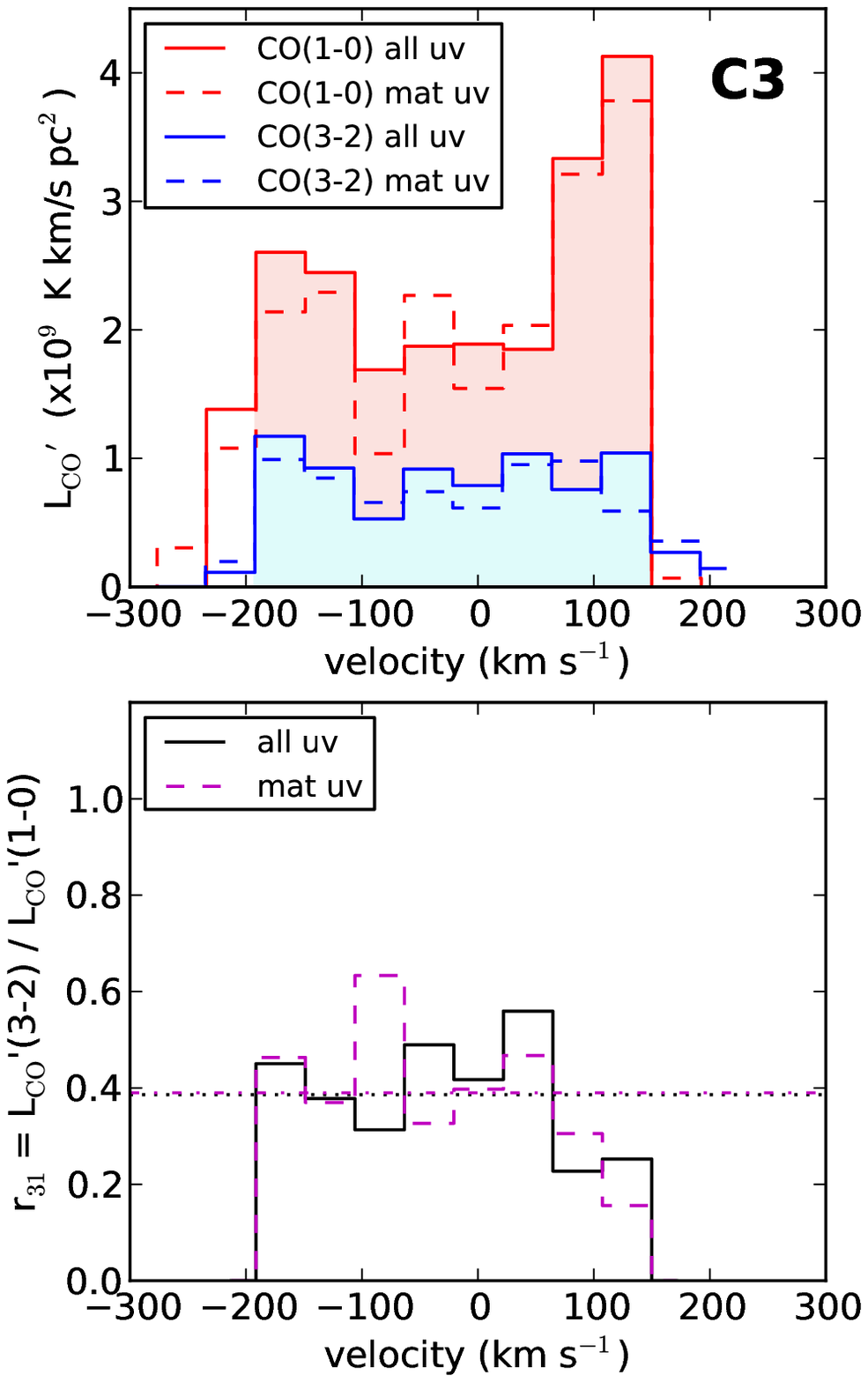}
\end{minipage}
\begin{minipage}[h]{0.32\linewidth}
\includegraphics[width=\linewidth]{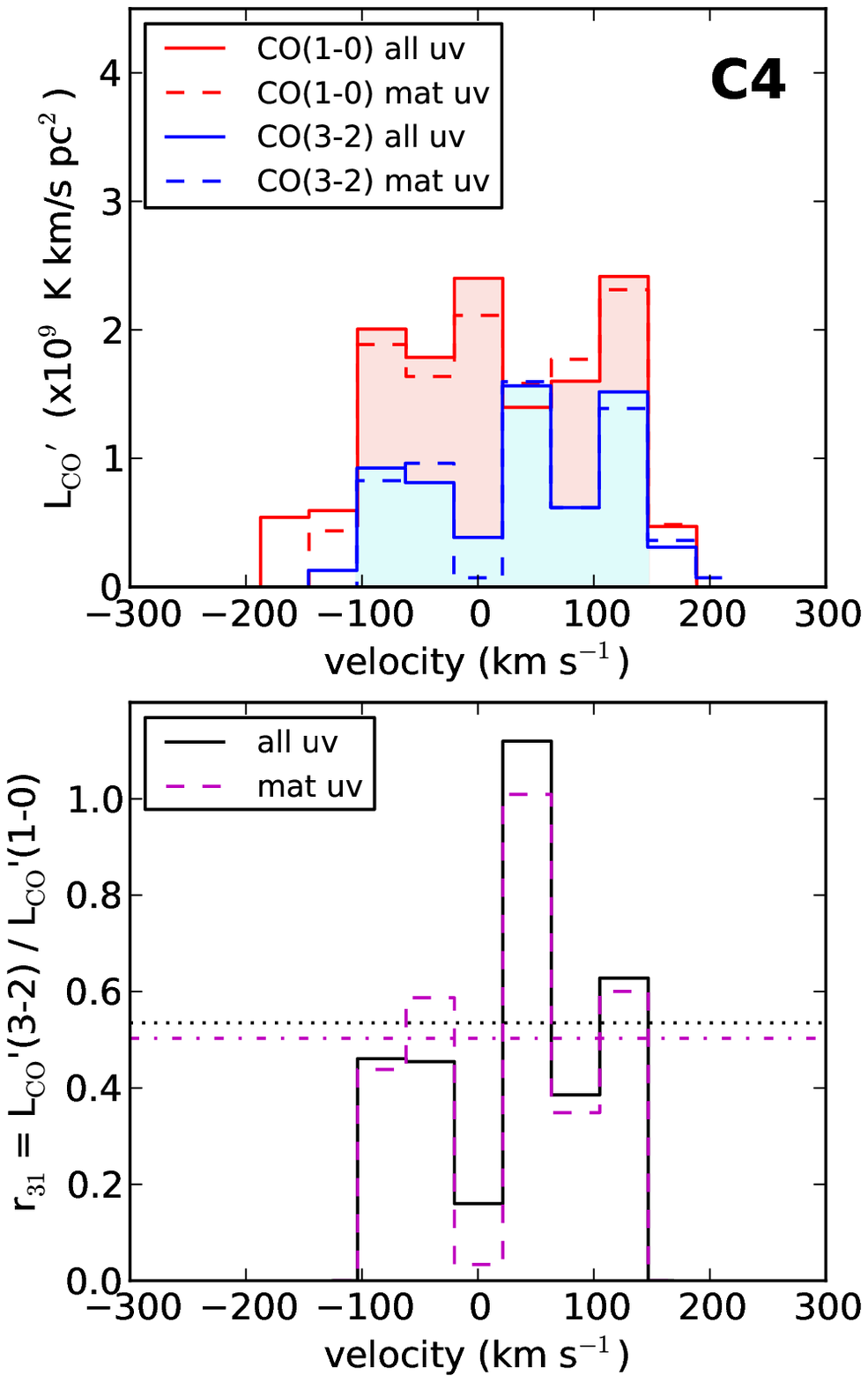}
\end{minipage}
\caption{$L_\mathrm{CO}'$ and $r_{31}$ profiles for the three detected sources. 
The top panels show the $L_\mathrm{CO}'$ profiles for the CO\jone (red) and CO\jthree (blue) transitions 
using all $uv$ data (all $uv$, solid lines) and matched $uv$ data (mat $uv$, dashed lines). 
The shading under each profile indicates which velocity channels are considered part of the source emission for each transition.
The $r_{31}$ profile is plotted in the bottom panel for both $uv$ selections (all $uv$ in solid black, matched $uv$ in dashed magenta). 
The average $r_{31}$ value for each profile is plotted as a horizontal black dotted (all $uv$ data) or 
magenta dash-dotted (matched $uv$ data) line.}
\label{r31profiles}
\end{figure*}

Next, to investigate the effect of the different $uv$ sampling of the two lines, 
we re-create the images using only data at $uv$ distances
present in both the 1 and 3 mm data (matched $uv$ data): 8-44 k$\lambda$, 
which corresponds to spatial scales of $\approx2.5$\arcsec-13\arcsec.
Further, we force all the images
to have a standard pixel size (0.99\arcsec for all except the CO\jthree line for C2 and C3 using all $uv$ data, 
for which we use 0.33\arcsec pixels since these images have higher resolution).
Figures \ref{C2fig} - \ref{C4fig} in Appendix \ref{sec:mommaps} show
images and spectra calculated within the standard source regions 
(discussed in Section \ref{sec:fluxest})
using all $uv$ data (top panel) and using matched $uv$ data (bottom panel).

This $uv$-distance restriction makes the beam sizes more similar, but not necessarily
the same due to different sampling within the allowed $uv$-distance range. 
Table \ref{tab:uvselfluxes} compares the measured fluxes and beam sizes using all $uv$ data
and matched $uv$ data. The change in the measured flux is $\lesssim 6\%$ for sources C2 and C4,
which is less than the expected measurement error in the total flux. However, the change in flux is more
significant for source C3 in both lines. For the CO\jone line, the matched 
$uv$-coverage excludes the short $uv$ spacings, which are sensitive to large-scale structure. 
Therefore, the presence of extended emission in source C3 would explain the decrease in flux
in the matched $uv$ data. This is supported by the radial profile of the CO\jone cumulative flux in Figure \ref{fig:radprofile}
in Appendix \ref{sec:datreducandflux}, 
which shows the matched $uv$ data agreeing with the all $uv$ data at small radii and diverging at larger radii.
The CO\jthree line, on the other hand, looses long $uv$ spacings (sensitive to small-scale structure) in the matched $uv$
data. However, the $uv$-distance restriction excludes a sizeable fraction of the $uv$ data, which decreases the signal
to noise in the image so that less flux is recovered using the masking technique. A comparison of the unmasked 
CO\jthree line fluxes in the all $uv$ and matched $uv$ datasets for source C3 shows a drop of only 7.2\%, which is
within the expected measurement error.

In the rest of our analysis, we consider the emission measured in both cases of $uv$ coverage.

\subsection{Integrated $r_{31}$ Velocity Profiles}
\label{sec:carefulr31}

We now make a more careful measurement of $r_{31}$, with the CO\jone and CO\jthree flux measurements
as well-matched to one another as possible. For each $uv$ data selection (all $uv$, matched $uv$), we calculate
$r_{31}$ in each channel where flux is measured in both lines. 

Figure \ref{r31profiles} shows the integrated $L_\mathrm{CO}'$ and $r_{31}$ velocity profiles 
for each of the three significantly detected galaxies. $L_\mathrm{CO}'$ as a function of velocity is plotted
in the top panels for each transition (CO\jone in red, CO\jthree in blue), 
for all $uv$ data (solid lines) and matched $uv$ data (dashed lines). Source velocity channels
for each line (chosen by eye) are indicated by shading in the corresponding color. The bottom panels show
the $r_{31}$ profiles for all $uv$ data (solid black line, with the average value indicated by the horizontal black dotted line)
and the matched $uv$ data (dashed magenta line, with average value shown by the horizontal magenta dash-dotted line). 
For galaxy C2, we have excluded the velocity channels at $\pm86$ \kmssp from this analysis.
At $-86$ \kms, we do not detect the CO\jthree line despite a $3\sigma$ detection in the CO\jone line. 
At $+86$ \kms, the CO\jthree image suggests a weak ($2\sigma$) detection at the center, but the measured flux is 
dominated by a $3\sigma$ noise spike close to this emission. Therefore, we use only the central three velocity channels 
in the CO\jthree transition.

In galaxies C2 and C3, the $r_{31}$ profiles appear well behaved, remaining roughly constant in all velocity channels.
On the other hand, galaxy C4 shows significant deviations from a flat profile, with a very low $r_{31}$ at $\approx0$ \kmssp and
a very high value at $\approx40$ \kms. Due to the modest signal to noise ratio at which the CO\jthree line
is detected in each channel, we cannot draw any definitive conclusions, but note that
if the enhanced $r_{31}$ is real, it may indicate a region of enhanced gas excitation (at a velocity close to 0, 
it is likely to be physically near the center of the galaxy), such as a starbursting clump,
in which a higher $r_{31}$ would be expected.

For each galaxy, Table \ref{tab:finalr31} gives the number of velocity channels in which $r_{31}$ is calculated, 
the mean and standard deviation ($\sigma$) of $r_{31}$ for each $uv$ data selection and the
the average $r_{31}$ value for the two $uv$ data selections (galaxy mean), with the expected measurement error (meas. error).
We expect a 20\% measurement error in the flux of each line, in each channel (Section \ref{sec:fluxest}), which gives 
a 30\% error in $r_{31}$ in each channel, and a $30\%(N_\mathrm{ch})^{-1/2}$
error in the average $r_{31}$. This value is reported in the last column of Table \ref{tab:finalr31} for each galaxy. 
The bottom rows give the mean and standard deviation of the average $r_{31}$ values for the three sample galaxies.
In this paper, the standard deviation given is the square root of the unbiased sample variance.

\begin{table}[t]
\centering
\begin{tabular}{| r | c | c | c | c | c | c | c |}
\hline
Name & $N_\mathrm{ch}$ & \multicolumn{2}{c|}{All $uv$} & \multicolumn{2}{c|}{Matched $uv$} & Galaxy & Meas. \\
\cline{3-6}
 &  & Mean & $\sigma$ & Mean & $\sigma$ & Mean & Error\\
\hline
C2 & 3 & 0.45 & 0.06 & 0.47 & 0.14 & 0.46 & 0.08 \\
C3 & 8 & 0.39 & 0.12 & 0.39 & 0.14 & 0.39 & 0.04 \\
C4 & 6 & 0.54 & 0.32 & 0.50 & 0.32 & 0.52 & 0.06 \\
\Cline{1.05pt}{1-8}
\multicolumn{6}{|r|}{Sample Mean} & 0.46 &  \\
\multicolumn{6}{|r|}{Sample $\sigma$} & 0.07 & \\
\hline
\end{tabular}
\caption{Summary of EGNoG galaxy $r_{31}$ values. For each source, we give the
number of velocity channels in which $r_{31}$ is calculated ($N_\mathrm{ch}$), the mean
and standard deviation ($\sigma$) of $r_{31}$ using all $uv$ data and the matched $uv$ data, 
and the average $r_{31}$ value for each galaxy, with the expected measurement error. 
In the bottom rows we give the mean and standard deviation of $r_{31}$ in this sample of three galaxies.
The expected measurement error for the average $r_{31}$ in each galaxy is the expected measurement
error in each channel (30\%) divided by $\sqrt{N_\mathrm{ch}}$.}
\label{tab:finalr31}
\end{table}

The average values for each $uv$ selection are consistent with each other within each galaxy 
and roughly consistent across the three galaxies. 
While the $uv$ selection appears to have a small effect on the average $r_{31}$ value, 
it is not in a systematic direction. 
The standard deviation of $r_{31}$ we observe in galaxies C2 and C3 is consistent with the expected 
30\% measurement errors in each channel. The standard deviation for galaxy C4 is roughly twice the 
expected 30\% error due to the two outlier channels discussed above. 
Overall, in the three EGNoG galaxies discussed here, we find a mean $r_{31}$ of 0.46 with a 
standard deviation of 0.07. Since we estimate each line flux has systematic errors of up to $\approx30\%$ 
(see Section \ref{sec:fluxest}), the systematic errors in the ratio $r_{31}$ will be $\lesssim40\%$.

\subsection{Radial Dependence of $r_{31}$}
\label{sec:fittingr31}

Since $r_{31}$ traces the local excitation conditions of the molecular gas of a galaxy, it is expected
to vary within the disk of the galaxy. In fact, \citet{Dumke2001} found CO\jthree emission to be more
centrally concentrated than CO\jone emission in nearby galaxies, so that $r_{31}$ decreased with radius. 
To look for radial variation in $r_{31}$ in the EGNoG data, we cannot compare the emission maps directly 
due to the marginal resolution of the galaxies and the different $uv$ coverage of the two transitions. 
In order to disentangle the true emission distribution from the $uv$ sampling, we fit a two-dimensional Gaussian 
to the total intensity (moment 0) map of each CO transition using the MIRIAD program {\tt imfit}. 
This analysis is presented in detail in Appendix \ref{sec:raddepappendix}.

While the error bars are large, we do find the ratio of the deconvolved peak intensities (representative of
the conditions in the central region of the galaxy) to be systematically higher than the ratio of the total
fluxes. In our sample, we find $r_{31}(\mathrm{peak}) = 0.73\pm0.31$, with a standard deviation across the three galaxies
of 0.04. We compare this to the ratio calculated from the fit total fluxes: $0.42\pm0.11$, with a standard deviation of 0.04.
While the small standard deviations in our $r_{31}$ values show consistency between the three galaxies, 
these results are plagued by sizable uncertainties as a result of fitting data with only modest signal to noise
(note that our error estimates are conservative and may in fact be over-estimating the true error).
Therefore, we conclude (but not robustly) that $r_{31}$ is higher in the center of the EGNoG galaxies 
than in the molecular disk as a whole \citep[consistent with][]{Dumke2001}.

\section{Discussion}
\label{sec:discussion}

The observed $r_{31}$ is a function of many parameters, including the excitation temperature ($T_\mathrm{ex}$), 
optical depth ($\tau_\nu$) and filling factor of each line \citep[e.g.][]{Hurt1993}. 
For each line, the rest-frame intensity is 
\begin{eqnarray}
\label{eqn:lineintensity}
I_{\nu} & = & \eta_\mathrm{f} \left(1 - \exp^{-\tau_\nu}\right) \left[ B_\nu(T_\mathrm{ex}) - B_\nu(T_\mathrm{CMB})\right] \\
 & \equiv & B_\nu(T_\mathrm{b}) \equiv \frac{2k\nu^2 T_\mathrm{b,RJ}}{c^2} \nonumber
\end{eqnarray}
%\begin{equation}
%I_{\nu}  =  \eta_\mathrm{f} \left(1 - \exp^{-\tau_\nu}\right) \left[ B_\nu(T_\mathrm{ex}) - B_\nu(T_\mathrm{CMB})\right] 
%\end{equation}
where $B_\nu(T)$ is the specific intensity of a blackbody at temperature $T$, 
$T_\mathrm{b}$ is the brightness temperature of the line, $T_\mathrm{b,RJ}$ is the Rayleigh-Jeans definition
brightness temperature of the line and $\eta_\mathrm{f}$ is the efficiency of the coupling between the beam and source emission, 
which is determined by the beam size, the molecular gas disk size and the fraction of the disk that is emitting in the line (the filling factor). 
Note that Equation \ref{eqn:lineintensity} is in the rest frame. The observed line intensity would be redshifted so that the corresponding
observed brightness temperature (for both definitions) is reduced by a factor of $(1+z)$.

For equal velocity widths in both lines (as we have enforced in Sections \ref{sec:carefulr31} and \ref{sec:fittingr31}), $r_{31}$, as defined in Equations
\ref{LCOdef} and \ref{r31equ}, reduces to the ratio of the Rayleigh-Jeans brightness temperatures. Following Equation \ref{eqn:lineintensity}, 
we can write $r_{31}$ as
\begin{equation}
r_{31} = \frac{T_\mathrm{b,RJ}(3-2)}{T_\mathrm{b,RJ}(1-0)} = \frac{\left[\eta_\mathrm{RJ} \eta_\mathrm{f} \eta_\mathrm{\tau} \eta_\mathrm{CMB} T_\mathrm{ex}\right]_{(3-2)}}{\left[\eta_\mathrm{RJ} \eta_\mathrm{f} \eta_\mathrm{\tau} \eta_\mathrm{CMB} T_\mathrm{ex}\right]_{(1-0)}}
\end{equation}
where each term in Equation \ref{eqn:lineintensity} is represented as a correction factor $\eta$ so that we can express $r_{31}$
more directly in terms of the line excitation temperatures. Specifically, $\eta_\mathrm{CMB}$ gives the correction from $T_\mathrm{ex}$ 
to the effective excitation temperature ($T_\mathrm{ex,eff}$, where $B_\nu(T_\mathrm{ex,eff}) = B_\nu(T_\mathrm{ex}) - B_\nu(T_\mathrm{CMB})$)
due to the CMB term. 
At $z = 0.3$, the temperature of the cosmic microwave background ($T_\mathrm{CMB}$) is $\approx 3.5$ K, so that 
for $T_\mathrm{ex} =$ 10 - 30 K, $\eta_\mathrm{CMB} \approx $ 0.85 - 0.95 for the CO\jone line and 0.98 - 1.00 for the CO\jthree line.
The term accounting for the optical depth of each line is $\eta_\tau = (1 - e^{-\tau_\nu})$.
The filling factor term, $\eta_\mathrm{f}$ (defined above) is expected to be $< 1$ and may be different for the two lines. 
Finally, we represent the difference between $T_\mathrm{b}$ and $T_\mathrm{b,RJ}$ 
by $\eta_\mathrm{RJ}$, which, for $T_\mathrm{b} =$ 10 - 30 K, is 0.75 - 0.91 for the CO\jone line and 0.39 - 0.75 for the CO\jthree line.
Therefore, for $T_\mathrm{ex} =$ 10 - 30 K for each line, we expect 
\begin{eqnarray}
r_{31}  & = & (0.4 - 1.2) \frac{\left[\eta_\mathrm{f} \eta_\tau T_\mathrm{ex}\right]_{3-2}}{\left[\eta_\mathrm{f} \eta_\tau T_\mathrm{ex}\right]_{1-0}}
\end{eqnarray}

The excitation temperature of a given emission line will fall between the radiation temperature ($T_\mathrm{CMB}$)
and the kinetic temperature, depending on the density of the gas.
Where the molecular gas density is larger than the critical density (for 20 K gas, 
$n_\mathrm{crit} \approx10^3$ cm$^{-3}$ for CO\jone and $\approx10^5$ cm$^{-3}$ for CO\jthree), 
the excitation temperature is equal to the kinetic temperature of the gas.
Thus we see that optically thick gas with equal excitation temperature in both lines
will have $r_{31} \approx 1.0 \pm 0.5$ for $T_\mathrm{ex} \approx 10-30$ K (depending on the relative filling factors and optical depths of the two lines).
This is consistent with the general criterion that gas with $r_{31} < 1$ is sub-thermal and $r_{31} > 1$ is thermalized.
Note that deviation from $r_{31} = 1$ in thermalized gas is a strong function of excitation temperature (due to $\eta_\mathrm{CMB}$
and $\eta_\mathrm{RJ}$): higher excitation temperature
gas will have $r_{31}$ closer to unity when thermalized.
More generally, if one considers various optical depths (assuming local thermodynamic equilibrium (LTE) with equal excitation temperatures),
$r_{31}$ greater than unity is indicative of warm, optically thin gas and a ratio less than unity is indicative of optically thick gas \citep[e.g.][]{Meier2001}.

Therefore, while we can comment on whether the excitation of the CO $J_\mathrm{upper}$ state is likely to be sub-thermal or thermalized, 
determining the conditions of the gas more accurately requires modeling of multiple transitions (e.g. using Large Velocity Gradient (LVG) models).
Since one line ratio does not provide strong constraints, we do not perform LVG modeling as part of this work.
In the following sections we compare the $r_{31}$ values found in the EGNoG galaxies to previous studies at
low and high redshift and we discuss the implications of this work for the interpretation of CO measurements in intermediate and high redshift galaxies.

\subsection{Comparison with Previous Work}
\label{sec:compprev}

To date, the study of CO line ratios in intermediate and high redshift galaxies has been dominated by work on
extreme starbursting systems: SMGs and quasars.
The CO lines from \jone up to \jnine in quasars at $z\approx 2-4$ are well-fit by a single component of 
highly excited gas \citep{Riechers2006, Weiss2007, Riechers2011z2quasars}.
In contrast, while the high-J CO transitions in $z\approx2-4$ SMGs are fit by similar highly excited gas, 
recent observations of the CO\jone line in these systems reveal a diffuse, low-excitation component in addition
to the highly excited component \citep{Carilli2010, Riechers2011z3SMGs}, 
similar to what has been observed in local ULIRGS \citep{Papa2007, Greve2009}.
In contrast, the limited work on $z\approx1-2$ SFGs suggests low-excitation gas 
similar to the Milky Way \citep{Dannerbauer2009, Aravena2010}.
Since the EGNoG galaxies are SFGs (lying on the main sequence), 
we restrict the rest of this discussion to SFGs (normal star-forming galaxies) at low and high redshift.

\begin{figure}[t]
\centering
\includegraphics[width=3.5in]{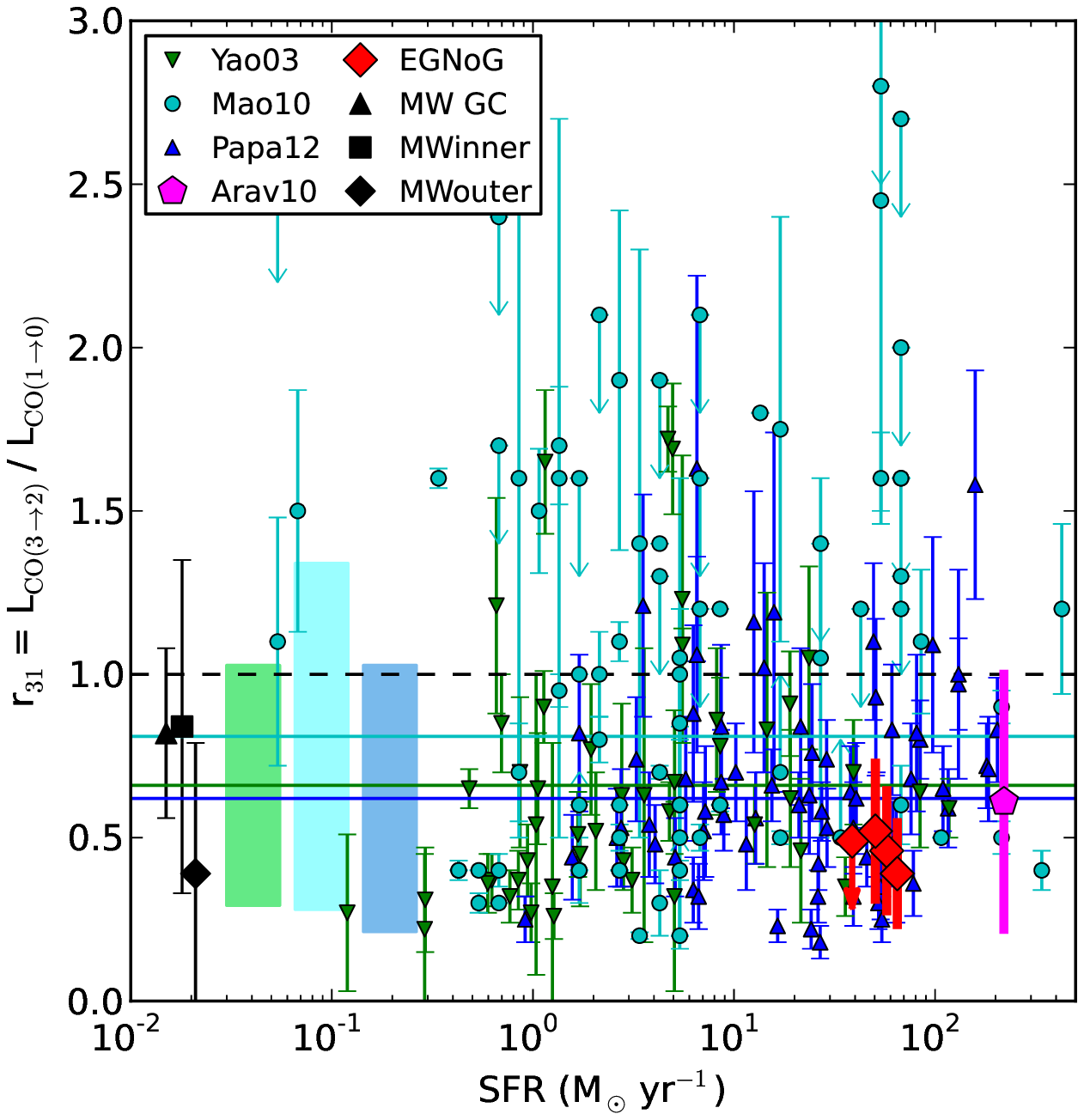}
\caption{Compilation of $r_{31}$ literature data. $r_{31}$ is plotted against approximate SFR 
($1.7\times10^{-10} L_\mathrm{FIR}$ when a SFR is not available) for the following datasets: 
\citet{Yao2003} (Yao03), \citet{Mao2010} (Mao10), \citet{Papa2011}
(Papa12), \citet{Aravena2010} (Arav10), this paper (EGNoG) and \citet{Fixsen1999} (MW GC, inner and outer).
Milky Way points are shown at the left side of the plot for clarity. 
The black horizontal dashed line 
shows $r_{31} = 1$ and the horizontal solid lines indicate the average values for Yao03, Mao10 and Papa11
(the corresponding shaded rectangles show the standard deviation around the average values). }
\label{fig:r31comp}
\end{figure}

To place the $r_{31}$ values we measure at $z\approx0.3$ in the context of previous work on normal
SFGs, Figure \ref{fig:r31comp}
shows $r_{31}$ versus approximate SFR for the EGNoG galaxies (red diamonds) and 
a compilation of literature data, which includes three large surveys of $r_{31}$ in nearby galaxies
\citep{Yao2003,Mao2010,Papa2011} as well as the study at $z=1.5$ \citep{Aravena2010}.
We plot $r_{31}$ from our matched analysis (Section \ref{sec:carefulr31}) for sources C2, C3 and C4
(values from Table \ref{tab:finalr31}, with 40\% error bars to indicate potential systematic errors)
and the upper limit for source C1 derived from the total line luminosities (Section \ref{sec:totr31}, Table \ref{tab:COprop}).
While SFRs are available for the present survey and \citet{Aravena2010}, the other three
surveys only provide total infrared luminosities ($L_\mathrm{IR}$) or far infrared luminosities ($L_\mathrm{FIR}$) calculated from 
IRAS (the Infrared Astronomical Satellite) fluxes. We use approximate
SFRs of $1.7\times10^{-10} L_\mathrm{FIR}$ M$_\odot$ yr$^{-1}$ ($L_\mathrm{FIR}$ in L$_\odot$; 
\citealt{Kennicutt1998ARAA}) and $L\mathrm{IR}/L_\mathrm{FIR} = 1.3$ \citep{GraciaCarpio2008} in Figure \ref{fig:r31comp}. 
Values for the Milky Way are plotted as well: the galactic center (MW GC), inner disk (MWinner),
and outer disk (MWouter) values, all in black, are taken from \citet{Fixsen1999} and plotted (slightly offset from one another) 
for illustrative purposes at the left side of the plot. The average $r_{31}$ values of the 
\citet{Yao2003}, \citet{Mao2010} and \citet{Papa2011}
datasets are indicated by the horizontal lines of the corresponding color, with the corresponding
shaded rectangles indicating the standard deviation around the average values. 
To guide the eye, the black dashed line shows $r_{31} = 1$.

Despite the large range of values observed, we find general agreement between our values, that of \citet{Aravena2010} and the average
values of the local surveys, with the EGNoG galaxies showing slightly lower $r_{31}$ values than the other studies. 
Note that the \citet{Mao2010} survey found an average value of 0.81 for the entire sample, 
but a lower average of 0.61 for galaxies classified as `normal', based on SFR surface density as indicated by the far infrared
luminosity and optical diameter.  
Further, the CO\jthree survey of local, low-SFR spirals by \citet{Wilson2012} find a lower value 
(using CO\jone luminosities from \citealt{Kuno2007}) of $r_{31}\approx0.2$, but note that the
authors have not made any correction for differences in the fraction of each galaxy mapped by the two surveys. 
Their $r_{31}$ value is at the low end of the spread of values plotted in Figure \ref{fig:r31comp}.
The range and average $r_{31}$ values for the points plotted (including
number of galaxies, redshift and SFR ranges) are given in Table \ref{tab:litdata}.

In order to compare our values to the literature data, we first discuss the nature of the observations presented.
\citet{Yao2003} observe 60 IR-luminous galaxies (most sources have $L_\mathrm{FIR} > 10^{10}$ L$_\odot$) 
selected from the SCUBA Local Universe Galaxy Survey (SLUGS).
Each line is measured with a single pointing of a single-dish telescope:
CO\jone at the Nobeyama Radio Observatory (NRO), CO\jthree at the James Clerk Maxwell Telescope (JCMT).
For both measurements, the beam size is $\approx 15\arcsec$, 
corresponding to physical sizes of $\approx0.5-13$ kpc for their sample
galaxies ($\approx0.5-5$ kpc for most of the sample). 

\citet{Mao2010} measure the CO\jthree line for 125 nearby galaxies of various types (e.g. normal, starburst, LIRG, ULIRG)
using the Heinrich Hertz Telescope (HHT; beam size $\approx 22\arcsec$). 
This beam size corresponds to a physical size of $\approx0.25-17$ kpc for the sample galaxies (1.7 kpc on average).
The CO\jone data for their sample were taken from the literature and therefore the beam size depends on the
telescope used. In their analysis, \citet{Mao2010} only use those galaxies for which they found IRAM 30m CO\jone
data (61), which have a beam size matching the HHT CO\jthree data. 
These measurements are plotted with error bars in Figure \ref{fig:r31comp}. 
Galaxies with CO\jone data from other sources are reported as upper or lower limits.

\begin{table}[t]
\centering
\begin{tabular}{| l | c | c | c | c | c |}
\hline
Dataset & N & Redshift  &  SFR$^1$ & \multicolumn{2}{c|}{$r_{31}$} \\
\cline{5-6}
 &  & Range  & (M$_\odot$ yr$^{-1}$) & Range & Mean \\
\hline
Milky Way & 3 &  &  & 0.4 - 0.9 & 0.68 \\
\hline
Mao 2010 & 61 & 0.0 - 0.04 & 0.03 - 200 & 0.2 - 1.9 & 0.81 \\
$\ \ $ normal & 7 &  &  & 0.3 - 1.5 & 0.61 \\
$\ \ $ starburst & 25 &  &  & 0.3 - 1.9 & 0.89 \\
\hline
Yao 2003 & 60 & 0.007 - 0.05 & 0.1 - 100 & 0.2 - 1.7 & 0.66 \\
\hline
Papa 2012 & 70 &  0.006 - 0.08 & 0.7 - 150 & 0.1 - 1.9 & 0.67 \\
\hline
EGNoG & 3 & 0.28 - 0.31 & 39 - 65 & 0.39 - 0.52 & 0.46 \\
\hline
Arav 2010 & 1 & 1.52 & 220 &  & 0.61 \\
\hline
\end{tabular}
\caption{Summary of $r_{31}$ from the literature. The range and mean values are given for the following datasets:
\citet{Fixsen1999} (Milky Way), \citet{Mao2010} (Mao 2010),
\citet{Yao2003} (Yao 2003), \citet{Papa2011} (Papa 2012), this paper (EGNoG) and \citet{Aravena2010} (Arav 2010). 
Columns 2 to 4 give the number of galaxies, redshift and SFR ranges for each dataset.
For Mao 2010, we also report the $r_{31}$ range and average for the `normal' and `starburst' subsets. \\
$^1$ SFR values are approximate (see Section \ref{sec:compprev}).}
\label{tab:litdata}
\end{table} 

\citet{Papa2011} examine a composite sample of 70 LIRGs in the nearby universe ($z \leq 0.1$)
spanning a wide range of morphologies.
They present new measurements of 36 galaxies and data from the literature for 34 more. 
The new measurements use the IRAM 30m telescope for CO\jone (beam size $\approx 22\arcsec$) 
and the JCMT for CO\jthree (beam size $\approx 14\arcsec$), and therefore are not observed with
the same beam size. The authors do not comment on matching beam sizes in the new measurements
or those from the literature.

The galaxy in which \citet{Aravena2010} measure $r_{31}$, BzK-21000 at $z=1.5$, was observed in the CO\jthree line with the 
Plateau de Bure Interferometer (PdBI) by \citet{Dannerbauer2009} and in the CO\jone line with the Very Large Array (VLA)
in C and D configurations \citep{Aravena2010}. In both cases, the source is unresolved or marginally resolved, so the measured
fluxes should represent the total emission of the galaxy.

The three samples of nearby galaxies are observed with single-dish instruments, typically sampling the inner portion of the molecular
gas disk of the observed galaxies. This is different from the $r_{31}$ measurements for the EGNoG galaxies (Table \ref{tab:finalr31}) 
and the galaxy at $z=1.5$ \citep{Aravena2010}, for which the emission from the entire gas disk is observed. 
We note that a gradient in $r_{31}$ has been observed in nearby galaxies \citep{Dumke2001}, 
with higher values measured in the centers. This can be inferred from the Milky Way
measurements \citep{Fixsen1999} as well. This effect may well account for the EGNoG
and \citet{Aravena2010} measurements appearing systematically lower than the average values reported by the surveys 
observing the central regions of nearby galaxies. 
Supporting this explanation, our analysis in Section \ref{sec:fittingr31} suggests $r_{31}$ is higher in the centers
of the EGNoG galaxies (more in line with the average values of the local surveys) compared to $r_{31}$ averaged over the entire disk. 
However, with such a large spread in values and inhomogeneous datasets, it is difficult to draw any robust conclusions on this point.

%\vspace{0.5cm}
\subsection{Implications}

We have measured $r_{31} = 0.46 \pm 0.07$ (with systematic errors of up to 40\%) in three galaxies at $z\approx0.3$,
suggestive of optically thick, sub-thermal gas. 
Despite being massive and highly star-forming (with SFRs of 50-65 M$_\odot$ yr$^{-1}$ 
and stellar masses of $\approx 2 \times 10^{11}$ M$_\odot$), the excitation of the gas in these galaxies is 
consistent with SFGs like local spirals, not star-bursting systems like ULIRGS, SMGs and quasars.
Since the EGNoG galaxies have been selected from the main sequence of star-forming galaxies at $z=0.3$,
our findings suggest that galaxies on the main sequence, over a range of star formation activity,
harbor sub-thermally excited gas. 
Therefore, we suggest that CO line ratios similar to those observed in local spiral galaxies 
are appropriate for main-sequence star-forming galaxies.

The extension of the EGNoG results to main-sequence star-forming galaxies
means that sub-thermal line ratios are appropriate for the $z\sim1-2$ SFGs in which 
\citet{Tacconi2010} and \citet{Daddi2010a} report high molecular gas fractions (20-80\%).
Tacconi et al. and Daddi et al. estimated the molecular gas mass associated with the observed 
$\mathrm{J}_\mathrm{upper}>1$ CO luminosity
assuming sub-thermal $r_\mathrm{J1}$ line ratios and a Milky Way-like $\alpha_\mathrm{CO}$
(\citealp{Daddi2010a} used a slightly smaller value for $\alpha_\mathrm{CO}$).
As the conversion factor is a complex problem, expected to vary as a function of gas excitation, density, 
metallicity and radiation field \citep{Shetty2011, Leroy2011}, the
results of this work do not directly inform the choice of conversion factor.
However, the $r_{31}$ line ratios of the EGNoG gas excitation sample 
support the assumption of a sub-thermal line ratio in these studies.
Specifically, \citet{Tacconi2010} use $r_{31} = 0.5$ to calculate molecular
gas masses, which is consistent with our results (\citealp{Daddi2010a} observed
the CO\jtwo line and assume a sub-thermal $r_{21}$).

\section{Conclusions}
\label{sec:conclusions}

This paper presents the gas excitation sample of the EGNoG survey.
We report robust detections of the CO\jthree and CO\jone lines
in three galaxies at $z\approx0.3$, and an upper limit for the fourth galaxy.
The average $r_{31}$ value for this sample is $0.46 \pm 0.07$, 
with systematic errors of up to 40\%.
This value is consistent with published $r_{31}$ values for main-sequence star-forming galaxies
at $z\approx0$ as well as the single measurement of at $z > 0.3$ \citep{Aravena2010}.
The sub-thermal excitation of the CO\jthree line suggests the 
excitation state of the molecular gas in these galaxies is similar to 
local spirals, and is not indicative of a starburst.
We conclude that the galaxies in our sample (and by extension, the main sequence 
galaxies at $z\sim1-2$ studied by \citealp{Tacconi2010} and \citealp{Daddi2010a}) 
harbor cold, optically thick molecular gas despite being massive and highly
star-forming.

\acknowledgements
The authors thank the anonymous referee for helpful comments and 
Eve Ostriker and Adam Leroy for useful discussions.
A. Bauermeister thanks Statia Cook, Dick Plambeck and Peter Williams 
for useful discussions on the reduction and analysis of the EGNoG data.
A. Bolatto wishes to acknowledge partial support from
grants CAREER NSF AST-0955836, NSF AST-1139998, as well as a Cottrell
Scholar award from the Research Corporation for Science Advancement.
We thank the OVRO/CARMA staff and the CARMA observers for their assistance in obtaining the data. 
Support for CARMA construction was derived from the Gordon and Betty Moore Foundation, 
the Kenneth T. and Eileen L. Norris Foundation, the James S. McDonnell Foundation, the 
Associates of the California Institute of Technology, the University of Chicago, the states of 
California, Illinois, and Maryland, and the National Science Foundation. Ongoing CARMA 
development and operations are supported by the National Science Foundation under a 
cooperative agreement, and by the CARMA partner universities.

\bibliographystyle{yahapj}
\bibliography{/Users/amber/Documents/research/myrefs}

\begin{thebibliography}{67}
\expandafter\ifx\csname natexlab\endcsname\relax\def\natexlab#1{#1}\fi

\bibitem[{{Abazajian} {et~al.}(2009){Abazajian}, {Adelman-McCarthy},
  {Ag{\"u}eros}, {Allam}, {Allende Prieto}, {An}, {Anderson}, {Anderson},
  {Annis}, {Bahcall}, \& et~al.}]{Abazajian2009}
{Abazajian}, K.~N., {Adelman-McCarthy}, J.~K., {Ag{\"u}eros}, M.~A., {et~al.}
  2009, \href{http://dx.doi.org/10.1088/0067-0049/182/2/543}{\apjs, 182, 543}

\bibitem[{{Aravena} {et~al.}(2010){Aravena}, {Carilli}, {Daddi}, {Wagg},
  {Walter}, {Riechers}, {Dannerbauer}, {Morrison}, {Stern}, \&
  {Krips}}]{Aravena2010}
{Aravena}, M., {Carilli}, C., {Daddi}, E., {et~al.} 2010,
  \href{http://dx.doi.org/10.1088/0004-637X/718/1/177}{\apj, 718, 177}

\bibitem[{{Baker} {et~al.}(2004){Baker}, {Tacconi}, {Genzel}, {Lehnert}, \&
  {Lutz}}]{Baker2004}
{Baker}, A.~J., {Tacconi}, L.~J., {Genzel}, R., {Lehnert}, M.~D., \& {Lutz}, D.
  2004, \href{http://dx.doi.org/10.1086/381798}{\apj, 604, 125}

\bibitem[{{Baldwin} {et~al.}(1981){Baldwin}, {Phillips}, \&
  {Terlevich}}]{BPT1981}
{Baldwin}, J.~A., {Phillips}, M.~M., \& {Terlevich}, R. 1981,
  \href{http://dx.doi.org/10.1086/130766}{\pasp, 93, 5}

\bibitem[{{Bayet} {et~al.}(2004){Bayet}, {Gerin}, {Phillips}, \&
  {Contursi}}]{Bayet2004}
{Bayet}, E., {Gerin}, M., {Phillips}, T.~G., \& {Contursi}, A. 2004,
  \href{http://dx.doi.org/10.1051/0004-6361:20035614}{\aap, 427, 45}

\bibitem[{{Bell} {et~al.}(2005){Bell}, {Papovich}, {Wolf}, {Le Floc'h},
  {Caldwell}, {Barden}, {Egami}, {McIntosh}, {Meisenheimer},
  {P{\'e}rez-Gonz{\'a}lez}, {Rieke}, {Rieke}, {Rigby}, \& {Rix}}]{Bell2005}
{Bell}, E.~F., {Papovich}, C., {Wolf}, C., {et~al.} 2005,
  \href{http://dx.doi.org/10.1086/429552}{\apj, 625, 23}

\bibitem[{{Bouch{\'e}} {et~al.}(2010){Bouch{\'e}}, {Dekel}, {Genzel}, {Genel},
  {Cresci}, {F{\"o}rster Schreiber}, {Shapiro}, {Davies}, \&
  {Tacconi}}]{Bouche2010}
{Bouch{\'e}}, N., {Dekel}, A., {Genzel}, R., {et~al.} 2010,
  \href{http://dx.doi.org/10.1088/0004-637X/718/2/1001}{\apj, 718, 1001}

\bibitem[{{Briggs}(1995)}]{Briggs1995}
{Briggs}, D.~S. 1995, in Bulletin of the American Astronomical Society,
  Vol.~27, American Astronomical Society Meeting Abstracts, \#112.02

\bibitem[{{Brinchmann} {et~al.}(2004){Brinchmann}, {Charlot}, {White},
  {Tremonti}, {Kauffmann}, {Heckman}, \& {Brinkmann}}]{Brinchmann2004}
{Brinchmann}, J., {Charlot}, S., {White}, S.~D.~M., {et~al.} 2004,
  \href{http://dx.doi.org/10.1111/j.1365-2966.2004.07881.x}{\mnras, 351, 1151}

\bibitem[{{Carilli} {et~al.}(2010){Carilli}, {Daddi}, {Riechers}, {Walter},
  {Weiss}, {Dannerbauer}, {Morrison}, {Wagg}, {Dav{\'e}}, {Elbaz}, {Stern},
  {Dickinson}, {Krips}, \& {Aravena}}]{Carilli2010}
{Carilli}, C.~L., {Daddi}, E., {Riechers}, D., {et~al.} 2010,
  \href{http://dx.doi.org/10.1088/0004-637X/714/2/1407}{\apj, 714, 1407}

\bibitem[{{Combes} {et~al.}(1994){Combes}, {Prugniel}, {Rampazzo}, \&
  {Sulentic}}]{Combes1994}
{Combes}, F., {Prugniel}, P., {Rampazzo}, R., \& {Sulentic}, J.~W. 1994, \aap,
  281, 725

\bibitem[{{Daddi} {et~al.}(2007){Daddi}, {Dickinson}, {Morrison}, {Chary},
  {Cimatti}, {Elbaz}, {Frayer}, {Renzini}, {Pope}, {Alexander}, {Bauer},
  {Giavalisco}, {Huynh}, {Kurk}, \& {Mignoli}}]{Daddi2007}
{Daddi}, E., {Dickinson}, M., {Morrison}, G., {et~al.} 2007,
  \href{http://dx.doi.org/10.1086/521818}{\apj, 670, 156}

\bibitem[{{Daddi} {et~al.}(2010){Daddi}, {Bournaud}, {Walter}, {Dannerbauer},
  {Carilli}, {Dickinson}, {Elbaz}, {Morrison}, {Riechers}, {Onodera}, {Salmi},
  {Krips}, \& {Stern}}]{Daddi2010a}
{Daddi}, E., {Bournaud}, F., {Walter}, F., {et~al.} 2010,
  \href{http://dx.doi.org/10.1088/0004-637X/713/1/686}{\apj, 713, 686}

\bibitem[{{Dame}(2011)}]{Dame2011}
{Dame}, T.~M. 2011, ArXiv e-prints,
  \href{http://arxiv.org/abs/1101.1499}{{\ttfamily arXiv:1101.1499
  [astro-ph.IM]}}

\bibitem[{{Dannerbauer} {et~al.}(2009){Dannerbauer}, {Daddi}, {Riechers},
  {Walter}, {Carilli}, {Dickinson}, {Elbaz}, \& {Morrison}}]{Dannerbauer2009}
{Dannerbauer}, H., {Daddi}, E., {Riechers}, D.~A., {et~al.} 2009,
  \href{http://dx.doi.org/10.1088/0004-637X/698/2/L178}{\apjl, 698, L178}

\bibitem[{{Downes} \& {Solomon}(1998)}]{DownSol1998}
{Downes}, D., \& {Solomon}, P.~M. 1998,
  \href{http://dx.doi.org/10.1086/306339}{\apj, 507, 615}

\bibitem[{{Dumke} {et~al.}(2001){Dumke}, {Nieten}, {Thuma}, {Wielebinski}, \&
  {Walsh}}]{Dumke2001}
{Dumke}, M., {Nieten}, C., {Thuma}, G., {Wielebinski}, R., \& {Walsh}, W. 2001,
  \href{http://dx.doi.org/10.1051/0004-6361:20010670}{\aap, 373, 853}

\bibitem[{{Dutton} {et~al.}(2010){Dutton}, {van den Bosch}, \&
  {Dekel}}]{Dutton2010}
{Dutton}, A.~A., {van den Bosch}, F.~C., \& {Dekel}, A. 2010,
  \href{http://dx.doi.org/10.1111/j.1365-2966.2010.16620.x}{\mnras, 405, 1690}

\bibitem[{{Elbaz} {et~al.}(2007){Elbaz}, {Daddi}, {Le Borgne}, {Dickinson},
  {Alexander}, {Chary}, {Starck}, {Brandt}, {Kitzbichler}, {MacDonald},
  {Nonino}, {Popesso}, {Stern}, \& {Vanzella}}]{Elbaz2007}
{Elbaz}, D., {Daddi}, E., {Le Borgne}, D., {et~al.} 2007,
  \href{http://dx.doi.org/10.1051/0004-6361:20077525}{\aap, 468, 33}

\bibitem[{{Elbaz} {et~al.}(2011){Elbaz}, {Dickinson}, {Hwang},
  {D{\'{\i}}az-Santos}, {Magdis}, {Magnelli}, {Le Borgne}, {Galliano},
  {Pannella}, {Chanial}, {Armus}, {Charmandaris}, {Daddi}, {Aussel}, {Popesso},
  {Kartaltepe}, {Altieri}, {Valtchanov}, {Coia}, {Dannerbauer}, {Dasyra},
  {Leiton}, {Mazzarella}, {Alexander}, {Buat}, {Burgarella}, {Chary}, {Gilli},
  {Ivison}, {Juneau}, {Le Floc'h}, {Lutz}, {Morrison}, {Mullaney}, {Murphy},
  {Pope}, {Scott}, {Brodwin}, {Calzetti}, {Cesarsky}, {Charlot}, {Dole},
  {Eisenhardt}, {Ferguson}, {F{\"o}rster Schreiber}, {Frayer}, {Giavalisco},
  {Huynh}, {Koekemoer}, {Papovich}, {Reddy}, {Surace}, {Teplitz}, {Yun}, \&
  {Wilson}}]{Elbaz2011}
{Elbaz}, D., {Dickinson}, M., {Hwang}, H.~S., {et~al.} 2011,
  \href{http://dx.doi.org/10.1051/0004-6361/201117239}{\aap, 533, A119}

\bibitem[{{Engel} {et~al.}(2010){Engel}, {Tacconi}, {Davies}, {Neri}, {Smail},
  {Chapman}, {Genzel}, {Cox}, {Greve}, {Ivison}, {Blain}, {Bertoldi}, \&
  {Omont}}]{Engel2010}
{Engel}, H., {Tacconi}, L.~J., {Davies}, R.~I., {et~al.} 2010,
  \href{http://dx.doi.org/10.1088/0004-637X/724/1/233}{\apj, 724, 233}

\bibitem[{{Fixsen} {et~al.}(1999){Fixsen}, {Bennett}, \& {Mather}}]{Fixsen1999}
{Fixsen}, D.~J., {Bennett}, C.~L., \& {Mather}, J.~C. 1999,
  \href{http://dx.doi.org/10.1086/307962}{\apj, 526, 207}

\bibitem[{{Graci{\'a}-Carpio} {et~al.}(2008){Graci{\'a}-Carpio},
  {Garc{\'{\i}}a-Burillo}, {Planesas}, {Fuente}, \& {Usero}}]{GraciaCarpio2008}
{Graci{\'a}-Carpio}, J., {Garc{\'{\i}}a-Burillo}, S., {Planesas}, P., {Fuente},
  A., \& {Usero}, A. 2008,
  \href{http://dx.doi.org/10.1051/0004-6361:20078223}{\aap, 479, 703}

\bibitem[{{Greve} {et~al.}(2009){Greve}, {Papadopoulos}, {Gao}, \&
  {Radford}}]{Greve2009}
{Greve}, T.~R., {Papadopoulos}, P.~P., {Gao}, Y., \& {Radford}, S.~J.~E. 2009,
  \href{http://dx.doi.org/10.1088/0004-637X/692/2/1432}{\apj, 692, 1432}

\bibitem[{{Helfer} {et~al.}(2003){Helfer}, {Thornley}, {Regan}, {Wong},
  {Sheth}, {Vogel}, {Blitz}, \& {Bock}}]{Helfer2003}
{Helfer}, T.~T., {Thornley}, M.~D., {Regan}, M.~W., {et~al.} 2003,
  \href{http://dx.doi.org/10.1086/346076}{\apjs, 145, 259}

\bibitem[{{Hurt} {et~al.}(1993){Hurt}, {Turner}, {Ho}, \& {Martin}}]{Hurt1993}
{Hurt}, R.~L., {Turner}, J.~L., {Ho}, P.~T.~P., \& {Martin}, R.~N. 1993,
  \href{http://dx.doi.org/10.1086/172313}{\apj, 404, 602}

\bibitem[{{Karim} {et~al.}(2011){Karim}, {Schinnerer},
  {Mart{\'{\i}}nez-Sansigre}, {Sargent}, {van der Wel}, {Rix}, {Ilbert},
  {Smol{\v c}i{\'c}}, {Carilli}, {Pannella}, {Koekemoer}, {Bell}, \&
  {Salvato}}]{Karim2011}
{Karim}, A., {Schinnerer}, E., {Mart{\'{\i}}nez-Sansigre}, A., {et~al.} 2011,
  \href{http://dx.doi.org/10.1088/0004-637X/730/2/61}{\apj, 730, 61}

\bibitem[{{Kauffmann} {et~al.}(2003){Kauffmann}, {Heckman}, {Tremonti},
  {Brinchmann}, {Charlot}, {White}, {Ridgway}, {Brinkmann}, {Fukugita}, {Hall},
  {Ivezi{\'c}}, {Richards}, \& {Schneider}}]{Kauffmann2003}
{Kauffmann}, G., {Heckman}, T.~M., {Tremonti}, C., {et~al.} 2003,
  \href{http://dx.doi.org/10.1111/j.1365-2966.2003.07154.x}{\mnras, 346, 1055}

\bibitem[{{Kennicutt}(1998)}]{Kennicutt1998ARAA}
{Kennicutt}, Jr., R.~C. 1998,
  \href{http://dx.doi.org/10.1146/annurev.astro.36.1.189}{\araa, 36, 189}

\bibitem[{{Kuno} {et~al.}(2007){Kuno}, {Sato}, {Nakanishi}, {Hirota}, {Tosaki},
  {Shioya}, {Sorai}, {Nakai}, {Nishiyama}, \& {Vila-Vilar{\'o}}}]{Kuno2007}
{Kuno}, N., {Sato}, N., {Nakanishi}, H., {et~al.} 2007, \pasj, 59, 117

\bibitem[{{Leroy} {et~al.}(2009){Leroy}, {Walter}, {Bigiel}, {Usero}, {Weiss},
  {Brinks}, {de Blok}, {Kennicutt}, {Schuster}, {Kramer}, {Wiesemeyer}, \&
  {Roussel}}]{Leroy2009}
{Leroy}, A.~K., {Walter}, F., {Bigiel}, F., {et~al.} 2009,
  \href{http://dx.doi.org/10.1088/0004-6256/137/6/4670}{\aj, 137, 4670}

\bibitem[{{Leroy} {et~al.}(2011){Leroy}, {Bolatto}, {Gordon}, {Sandstrom},
  {Gratier}, {Rosolowsky}, {Engelbracht}, {Mizuno}, {Corbelli}, {Fukui}, \&
  {Kawamura}}]{Leroy2011}
{Leroy}, A.~K., {Bolatto}, A., {Gordon}, K., {et~al.} 2011,
  \href{http://dx.doi.org/10.1088/0004-637X/737/1/12}{\apj, 737, 12}

\bibitem[{{Mao} {et~al.}(2010){Mao}, {Schulz}, {Henkel}, {Mauersberger},
  {Muders}, \& {Dinh-V-Trung}}]{Mao2010}
{Mao}, R.-Q., {Schulz}, A., {Henkel}, C., {et~al.} 2010,
  \href{http://dx.doi.org/10.1088/0004-637X/724/2/1336}{\apj, 724, 1336}

\bibitem[{{Mauersberger} {et~al.}(1999){Mauersberger}, {Henkel}, {Walsh}, \&
  {Schulz}}]{Mauersberger1999}
{Mauersberger}, R., {Henkel}, C., {Walsh}, W., \& {Schulz}, A. 1999, \aap, 341,
  256

\bibitem[{{Meier} {et~al.}(2001){Meier}, {Turner}, {Crosthwaite}, \&
  {Beck}}]{Meier2001}
{Meier}, D.~S., {Turner}, J.~L., {Crosthwaite}, L.~P., \& {Beck}, S.~C. 2001,
  \href{http://dx.doi.org/10.1086/318782}{\aj, 121, 740}

\bibitem[{{Noeske} {et~al.}(2007){Noeske}, {Faber}, {Weiner}, {Koo}, {Primack},
  {Dekel}, {Papovich}, {Conselice}, {Le Floc'h}, {Rieke}, {Coil}, {Lotz},
  {Somerville}, \& {Bundy}}]{Noeske2007}
{Noeske}, K.~G., {Faber}, S.~M., {Weiner}, B.~J., {et~al.} 2007,
  \href{http://dx.doi.org/10.1086/517927}{\apjl, 660, L47}

\bibitem[{{Pannella} {et~al.}(2009){Pannella}, {Carilli}, {Daddi}, {McCracken},
  {Owen}, {Renzini}, {Strazzullo}, {Civano}, {Koekemoer}, {Schinnerer},
  {Scoville}, {Smol{\v c}i{\'c}}, {Taniguchi}, {Aussel}, {Kneib}, {Ilbert},
  {Mellier}, {Salvato}, {Thompson}, \& {Willott}}]{Pannella2009}
{Pannella}, M., {Carilli}, C.~L., {Daddi}, E., {et~al.} 2009,
  \href{http://dx.doi.org/10.1088/0004-637X/698/2/L116}{\apjl, 698, L116}

\bibitem[{{Papadopoulos} {et~al.}(2007){Papadopoulos}, {Isaak}, \& {van der
  Werf}}]{Papa2007}
{Papadopoulos}, P.~P., {Isaak}, K.~G., \& {van der Werf}, P.~P. 2007,
  \href{http://dx.doi.org/10.1086/520671}{\apj, 668, 815}

\bibitem[{{Papadopoulos} {et~al.}(2012){Papadopoulos}, {van der Werf},
  {Xilouris}, {Isaak}, {Gao}, \& {M{\"u}hle}}]{Papa2011}
{Papadopoulos}, P.~P., {van der Werf}, P.~P., {Xilouris}, E.~M., {et~al.} 2012,
  \href{http://dx.doi.org/10.1111/j.1365-2966.2012.21001.x}{\mnras, 426, 2601}

\bibitem[{{Polychroni} {et~al.}(2012){Polychroni}, {Moore}, \&
  {Allsopp}}]{Polychroni2012}
{Polychroni}, D., {Moore}, T.~J.~T., \& {Allsopp}, J. 2012,
  \href{http://dx.doi.org/10.1111/j.1365-2966.2012.20803.x}{\mnras, 422, 2992}

\bibitem[{{Pound} \& {Teuben}(2012)}]{PoundTeuben2012}
{Pound}, M.~W., \& {Teuben}, P. 2012, in Astronomical Society of the Pacific
  Conference Series, Vol. 461, Astronomical Data Analysis Software and Systems
  XXI, ed. P.~{Ballester}, D.~{Egret}, \& N.~P.~F. {Lorente}, 565

\bibitem[{{Rahman} {et~al.}(2012){Rahman}, {Bolatto}, {Xue}, {Wong}, {Leroy},
  {Walter}, {Bigiel}, {Rosolowsky}, {Fisher}, {Vogel}, {Blitz}, {West}, \&
  {Ott}}]{Rahman2012}
{Rahman}, N., {Bolatto}, A.~D., {Xue}, R., {et~al.} 2012,
  \href{http://dx.doi.org/10.1088/0004-637X/745/2/183}{\apj, 745, 183}

\bibitem[{{Regan} {et~al.}(2001){Regan}, {Thornley}, {Helfer}, {Sheth}, {Wong},
  {Vogel}, {Blitz}, \& {Bock}}]{Regan2001}
{Regan}, M.~W., {Thornley}, M.~D., {Helfer}, T.~T., {et~al.} 2001,
  \href{http://dx.doi.org/10.1086/323221}{\apj, 561, 218}

\bibitem[{{Rickard} \& {Blitz}(1985)}]{RickardBlitz1985}
{Rickard}, L.~J., \& {Blitz}, L. 1985,
  \href{http://dx.doi.org/10.1086/184472}{\apjl, 292, L57}

\bibitem[{{Riechers}(2011)}]{Riechers2011z2quasars}
{Riechers}, D.~A. 2011,
  \href{http://dx.doi.org/10.1088/0004-637X/730/2/108}{\apj, 730, 108}

\bibitem[{{Riechers} {et~al.}(2011){Riechers}, {Hodge}, {Walter}, {Carilli}, \&
  {Bertoldi}}]{Riechers2011z3SMGs}
{Riechers}, D.~A., {Hodge}, J., {Walter}, F., {Carilli}, C.~L., \& {Bertoldi},
  F. 2011, \href{http://dx.doi.org/10.1088/2041-8205/739/1/L31}{\apjl, 739,
  L31}

\bibitem[{{Riechers} {et~al.}(2006){Riechers}, {Walter}, {Carilli}, {Knudsen},
  {Lo}, {Benford}, {Staguhn}, {Hunter}, {Bertoldi}, {Henkel}, {Menten},
  {Weiss}, {Yun}, \& {Scoville}}]{Riechers2006}
{Riechers}, D.~A., {Walter}, F., {Carilli}, C.~L., {et~al.} 2006,
  \href{http://dx.doi.org/10.1086/507014}{\apj, 650, 604}

\bibitem[{{Rodighiero} {et~al.}(2011){Rodighiero}, {Daddi}, {Baronchelli},
  {Cimatti}, {Renzini}, {Aussel}, {Popesso}, {Lutz}, {Andreani}, {Berta},
  {Cava}, {Elbaz}, {Feltre}, {Fontana}, {F{\"o}rster Schreiber},
  {Franceschini}, {Genzel}, {Grazian}, {Gruppioni}, {Ilbert}, {Le Floch},
  {Magdis}, {Magliocchetti}, {Magnelli}, {Maiolino}, {McCracken}, {Nordon},
  {Poglitsch}, {Santini}, {Pozzi}, {Riguccini}, {Tacconi}, {Wuyts}, \&
  {Zamorani}}]{Rodighiero2011}
{Rodighiero}, G., {Daddi}, E., {Baronchelli}, I., {et~al.} 2011,
  \href{http://dx.doi.org/10.1088/2041-8205/739/2/L40}{\apjl, 739, L40}

\bibitem[{{Sanders} {et~al.}(1986){Sanders}, {Scoville}, {Young}, {Soifer},
  {Schloerb}, {Rice}, \& {Danielson}}]{Sanders1986}
{Sanders}, D.~B., {Scoville}, N.~Z., {Young}, J.~S., {et~al.} 1986,
  \href{http://dx.doi.org/10.1086/184682}{\apjl, 305, L45}

\bibitem[{{Sault} {et~al.}(2011){Sault}, {Teuben}, \& {Wright}}]{MIRIAD}
{Sault}, R.~J., {Teuben}, P.~J., \& {Wright}, M.~C.~H. 2011, in Astrophysics
  Source Code Library, record ascl:1106.007, 6007

\bibitem[{{Sheth} {et~al.}(2004){Sheth}, {Blain}, {Kneib}, {Frayer}, {van der
  Werf}, \& {Knudsen}}]{Sheth2004}
{Sheth}, K., {Blain}, A.~W., {Kneib}, J.-P., {et~al.} 2004,
  \href{http://dx.doi.org/10.1086/425308}{\apjl, 614, L5}

\bibitem[{{Shetty} {et~al.}(2011){Shetty}, {Glover}, {Dullemond}, {Ostriker},
  {Harris}, \& {Klessen}}]{Shetty2011}
{Shetty}, R., {Glover}, S.~C., {Dullemond}, C.~P., {et~al.} 2011,
  \href{http://dx.doi.org/10.1111/j.1365-2966.2011.18937.x}{\mnras, 415, 3253}

\bibitem[{{Solomon} {et~al.}(1997){Solomon}, {Downes}, {Radford}, \&
  {Barrett}}]{Solomon1997}
{Solomon}, P.~M., {Downes}, D., {Radford}, S.~J.~E., \& {Barrett}, J.~W. 1997,
  \href{http://dx.doi.org/10.1086/303765}{\apj, 478, 144}

\bibitem[{{Solomon} \& {Vanden Bout}(2005)}]{SolomonVandenBout2005}
{Solomon}, P.~M., \& {Vanden Bout}, P.~A. 2005,
  \href{http://dx.doi.org/10.1146/annurev.astro.43.051804.102221}{\araa, 43,
  677}

\bibitem[{{Strauss} {et~al.}(2002){Strauss}, {Weinberg}, {Lupton}, {Narayanan},
  {Annis}, {Bernardi}, {Blanton}, {Burles}, {Connolly}, {Dalcanton}, {Doi},
  {Eisenstein}, {Frieman}, {Fukugita}, {Gunn}, {Ivezi{\'c}}, {Kent}, {Kim},
  {Knapp}, {Kron}, {Munn}, {Newberg}, {Nichol}, {Okamura}, {Quinn}, {Richmond},
  {Schlegel}, {Shimasaku}, {SubbaRao}, {Szalay}, {Vanden Berk}, {Vogeley},
  {Yanny}, {Yasuda}, {York}, \& {Zehavi}}]{Strauss2002}
{Strauss}, M.~A., {Weinberg}, D.~H., {Lupton}, R.~H., {et~al.} 2002,
  \href{http://dx.doi.org/10.1086/342343}{\aj, 124, 1810}

\bibitem[{{Tacconi} {et~al.}(2008){Tacconi}, {Genzel}, {Smail}, {Neri},
  {Chapman}, {Ivison}, {Blain}, {Cox}, {Omont}, {Bertoldi}, {Greve},
  {F{\"o}rster Schreiber}, {Genel}, {Lutz}, {Swinbank}, {Shapley}, {Erb},
  {Cimatti}, {Daddi}, \& {Baker}}]{Tacconi2008}
{Tacconi}, L.~J., {Genzel}, R., {Smail}, I., {et~al.} 2008,
  \href{http://dx.doi.org/10.1086/587168}{\apj, 680, 246}

\bibitem[{{Tacconi} {et~al.}(2010){Tacconi}, {Genzel}, {Neri}, {Cox}, {Cooper},
  {Shapiro}, {Bolatto}, {Bouch{\'e}}, {Bournaud}, {Burkert}, {Combes},
  {Comerford}, {Davis}, {Schreiber}, {Garcia-Burillo}, {Gracia-Carpio}, {Lutz},
  {Naab}, {Omont}, {Shapley}, {Sternberg}, \& {Weiner}}]{Tacconi2010}
{Tacconi}, L.~J., {Genzel}, R., {Neri}, R., {et~al.} 2010,
  \href{http://dx.doi.org/10.1038/nature08773}{\nat, 463, 781}

\bibitem[{{Wang} {et~al.}(2006){Wang}, {Xia}, {Mao}, {Cao}, {Wu}, \&
  {Deng}}]{wang2006}
{Wang}, J.~L., {Xia}, X.~Y., {Mao}, S., {et~al.} 2006,
  \href{http://dx.doi.org/10.1086/506902}{\apj, 649, 722}

\bibitem[{{Wei{\ss}} {et~al.}(2007){Wei{\ss}}, {Downes}, {Neri}, {Walter},
  {Henkel}, {Wilner}, {Wagg}, \& {Wiklind}}]{Weiss2007}
{Wei{\ss}}, A., {Downes}, D., {Neri}, R., {et~al.} 2007,
  \href{http://dx.doi.org/10.1051/0004-6361:20066117}{\aap, 467, 955}

\bibitem[{{Wei{\ss}} {et~al.}(2005{\natexlab{a}}){Wei{\ss}}, {Downes},
  {Walter}, \& {Henkel}}]{Weiss2005b}
{Wei{\ss}}, A., {Downes}, D., {Walter}, F., \& {Henkel}, C. 2005{\natexlab{a}},
  \href{http://dx.doi.org/10.1051/0004-6361:200500166}{\aap, 440, L45}

\bibitem[{{Wei{\ss}} {et~al.}(2005{\natexlab{b}}){Wei{\ss}}, {Walter}, \&
  {Scoville}}]{Weiss2005a}
{Wei{\ss}}, A., {Walter}, F., \& {Scoville}, N.~Z. 2005{\natexlab{b}},
  \href{http://dx.doi.org/10.1051/0004-6361:20052667}{\aap, 438, 533}

\bibitem[{{Williams} {et~al.}(2012){Williams}, {Law}, \&
  {Bower}}]{miriadpython}
{Williams}, P.~K.~G., {Law}, C.~J., \& {Bower}, G.~C. 2012,
  \href{http://dx.doi.org/10.1086/666604}{\pasp, 124, 624}

\bibitem[{{Wilson} {et~al.}(2012){Wilson}, {Warren}, {Israel}, {Serjeant},
  {Attewell}, {Bendo}, {Butner}, {Chanial}, {Clements}, {Golding}, {Heesen},
  {Irwin}, {Leech}, {Matthews}, {M{\"u}hle}, {Mortier}, {Petitpas},
  {S{\'a}nchez-Gallego}, {Sinukoff}, {Shorten}, {Tan}, {Tilanus}, {Usero},
  {Vaccari}, {Wiegert}, {Zhu}, {Alexander}, {Alexander}, {Azimlu}, {Barmby},
  {Brar}, {Bridge}, {Brinks}, {Brooks}, {Coppin}, {C{\^o}t{\'e}},
  {C{\^o}t{\'e}}, {Courteau}, {Davies}, {Eales}, {Fich}, {Hudson}, {Hughes},
  {Ivison}, {Knapen}, {Page}, {Parkin}, {Rigopoulou}, {Rosolowsky}, {Seaquist},
  {Spekkens}, {Tanvir}, {van der Hulst}, {van der Werf}, {Vlahakis}, {Webb},
  {Weferling}, \& {White}}]{Wilson2012}
{Wilson}, C.~D., {Warren}, B.~E., {Israel}, F.~P., {et~al.} 2012,
  \href{http://dx.doi.org/10.1111/j.1365-2966.2012.21453.x}{\mnras, 424, 3050}

\bibitem[{{Wright} \& {Corder}(2008)}]{SKAmemo103}
{Wright}, M., \& {Corder}, S. 2008,
  \href{http://www.skatelescope.org/uploaded/57231_103_Memo_Wright.pdf}{SKA
  Memorandum Series, \#103}

\bibitem[{{Yao} {et~al.}(2003){Yao}, {Seaquist}, {Kuno}, \& {Dunne}}]{Yao2003}
{Yao}, L., {Seaquist}, E.~R., {Kuno}, N., \& {Dunne}, L. 2003,
  \href{http://dx.doi.org/10.1086/374333}{\apj, 588, 771}

\bibitem[{{York} {et~al.}(2000){York}, {Adelman}, {Anderson}, {Anderson},
  {Annis}, {Bahcall}, {Bakken}, {Barkhouser}, {Bastian}, {Berman}, {Boroski},
  {Bracker}, {Briegel}, {Briggs}, {Brinkmann}, {Brunner}, {Burles}, {Carey},
  {Carr}, {Castander}, {Chen}, {Colestock}, {Connolly}, {Crocker}, {Csabai},
  {Czarapata}, {Davis}, {Doi}, {Dombeck}, {Eisenstein}, {Ellman}, {Elms},
  {Evans}, {Fan}, {Federwitz}, {Fiscelli}, {Friedman}, {Frieman}, {Fukugita},
  {Gillespie}, {Gunn}, {Gurbani}, {de Haas}, {Haldeman}, {Harris}, {Hayes},
  {Heckman}, {Hennessy}, {Hindsley}, {Holm}, {Holmgren}, {Huang}, {Hull},
  {Husby}, {Ichikawa}, {Ichikawa}, {Ivezi{\'c}}, {Kent}, {Kim}, {Kinney},
  {Klaene}, {Kleinman}, {Kleinman}, {Knapp}, {Korienek}, {Kron}, {Kunszt},
  {Lamb}, {Lee}, {Leger}, {Limmongkol}, {Lindenmeyer}, {Long}, {Loomis},
  {Loveday}, {Lucinio}, {Lupton}, {MacKinnon}, {Mannery}, {Mantsch}, {Margon},
  {McGehee}, {McKay}, {Meiksin}, {Merelli}, {Monet}, {Munn}, {Narayanan},
  {Nash}, {Neilsen}, {Neswold}, {Newberg}, {Nichol}, {Nicinski}, {Nonino},
  {Okada}, {Okamura}, {Ostriker}, {Owen}, {Pauls}, {Peoples}, {Peterson},
  {Petravick}, {Pier}, {Pope}, {Pordes}, {Prosapio}, {Rechenmacher}, {Quinn},
  {Richards}, {Richmond}, {Rivetta}, {Rockosi}, {Ruthmansdorfer}, {Sandford},
  {Schlegel}, {Schneider}, {Sekiguchi}, {Sergey}, {Shimasaku}, {Siegmund},
  {Smee}, {Smith}, {Snedden}, {Stone}, {Stoughton}, {Strauss}, {Stubbs},
  {SubbaRao}, {Szalay}, {Szapudi}, {Szokoly}, {Thakar}, {Tremonti}, {Tucker},
  {Uomoto}, {Vanden Berk}, {Vogeley}, {Waddell}, {Wang}, {Watanabe},
  {Weinberg}, {Yanny}, {Yasuda}, \& {SDSS Collaboration}}]{York2000}
{York}, D.~G., {Adelman}, J., {Anderson}, Jr., J.~E., {et~al.} 2000,
  \href{http://dx.doi.org/10.1086/301513}{\aj, 120, 1579}

\bibitem[{{Young} {et~al.}(1995){Young}, {Xie}, {Tacconi}, {Knezek}, {Viscuso},
  {Tacconi-Garman}, {Scoville}, {Schneider}, {Schloerb}, {Lord}, {Lesser},
  {Kenney}, {Huang}, {Devereux}, {Claussen}, {Case}, {Carpenter}, {Berry}, \&
  {Allen}}]{Young1995}
{Young}, J.~S., {Xie}, S., {Tacconi}, L., {et~al.} 1995,
  \href{http://dx.doi.org/10.1086/192159}{\apjs, 98, 219}

\end{thebibliography}
%when ready, copy myrefs.bib to local dir and switch to that version:
%\bibliography{myrefs}

\begin{appendix}

\section{Data Reduction and Flux Measurement}
\label{sec:datreducandflux}
\subsection{Data Reduction}
Each dataset is reduced and calibrated as follows.
The data are flagged for antenna - antenna shadowing as well
as any other issues during the observation. The instrument bandpass is calibrated with
{\tt mfcal} on a bright passband calibrator. The time-dependent antenna gains (from atmospheric variation)
are derived by performing a {\tt selfcal} on the phase calibrator with an averaging interval of 18 minutes 
(the timescale of switching between the source and phase calibrator). For sources C1 and C2, 
the phase calibrator, 0854+201, is $\approx10\%$ polarized, which required additional steps in 
the reduction of the 3mm data (observed with linearly polarized feeds). This is
described in more detail in the full survey paper, Bauermeister et al. 2013b, {\it in preparation}.

For each dataset, the flux of the phase calibrator is set during the antenna
gain calibration in order to properly set the flux scale of the data. The flux of each phase
calibrator is assumed to be constant over timescales of weeks, and is 
therefore determined from the best datasets of the survey using
{\tt bootflux} on bandpass-calibrated, phase-only gain-calibrated data with Mars
or MWC349 as a primary flux calibrator. 
The brightness temperature of Mars is set by the CARMA system using the Caltech thermal
model of Mars (courtesy of Mark Gurwell), which includes seasonal variations in temperature
and can be accessed in MIRIAD using {\tt marstb}. This model gives brightness temperatures 
$\approx218$ K at 266 GHz and $\approx208-200$ K at 88 GHz for August-November 2011 observations, 
and $\approx201$ K at 266 GHz for April 2012 observations. 
For MWC349 we set the flux to 1.2 Jy at 88 GHz, the typical value from historical flux
monitoring at CARMA. The fluxes used for each phase calibrator are as follows:
for the 3mm data ($\approx 89$ GHz, August to November 2010), the flux of 0854+201 is set to 4 Jy (9\% linearly polarized), 
1357+193 0.8 Jy and 1310+323 1.7 Jy. For the 1mm data ($\approx 265$ GHz, August 2011 and April 2012)
the flux of 0854+201 is set to 2 Jy in August 2011 and 4 Jy in April 2012, 1224+213 0.6 Jy in April 2012 and 1310+323 0.6 Jy
in August 2011. 

Images are produced combining all fully calibrated datasets for each source. 
We used {\tt invert}, weighting the visibilities by the system temperature as well as using
a Briggs' visibility weighting robustness parameter \citep{Briggs1995} of 0.5. 
Since CARMA is an inhomogeneous array (these data use both the 10m and 6m dishes), 
we also use {\tt options=mosaic} in the {\tt invert} step in order to properly handle the three
different primary beams patterns (10m-10m, 10m-6m and 6m-6m). All observations are single-pointing.
The resulting images are primary-beam-corrected. In most of the sources discussed 
here, we map the 3mm data (CO\jone line) channel by channel, using the full spectral resolution
of 42 \kms. We match this resolution in the 1mm data (CO\jthree line) by averaging channels in sets of three.
The exception to this scheme is source C1, which has
emission spread over a very wide velocity range, requiring more channel averaging
to increase the signal to noise. 

We deconvolve each image with {\tt mossdi} (the mosaic version of {\tt clean}), 
cleaning down to the rms noise within a single channel, within a cleaning box selected by eye to include
only source emission. We clean only channels which contain visible source emission.
Cleaning down to a specified noise level is preferred to using a set number of clean iterations
due to the nature of the spectral line emission: some channels will contain more flux and
therefore require more clean iterations. In tests using a model source of known flux inserted
into real data (emission-free channels), we found a $1\sigma$ cutoff to best extract
the true source emission without overestimating the flux over a range of detection significance levels, with
a 10-30\% uncertainty in the recovered flux (depending on the significance of the signal).
The final clean images are produced by {\tt restor}, which convolves the clean component image with 
the clean beam (calculated by fitting a Gaussian to the combined mosaic beam given 
by {\tt mospsf}), and adds the residuals from the cleaning process. 

\begin{figure*}[t]
\centering
\begin{minipage}[h]{0.32\linewidth}
\includegraphics[width=\linewidth]{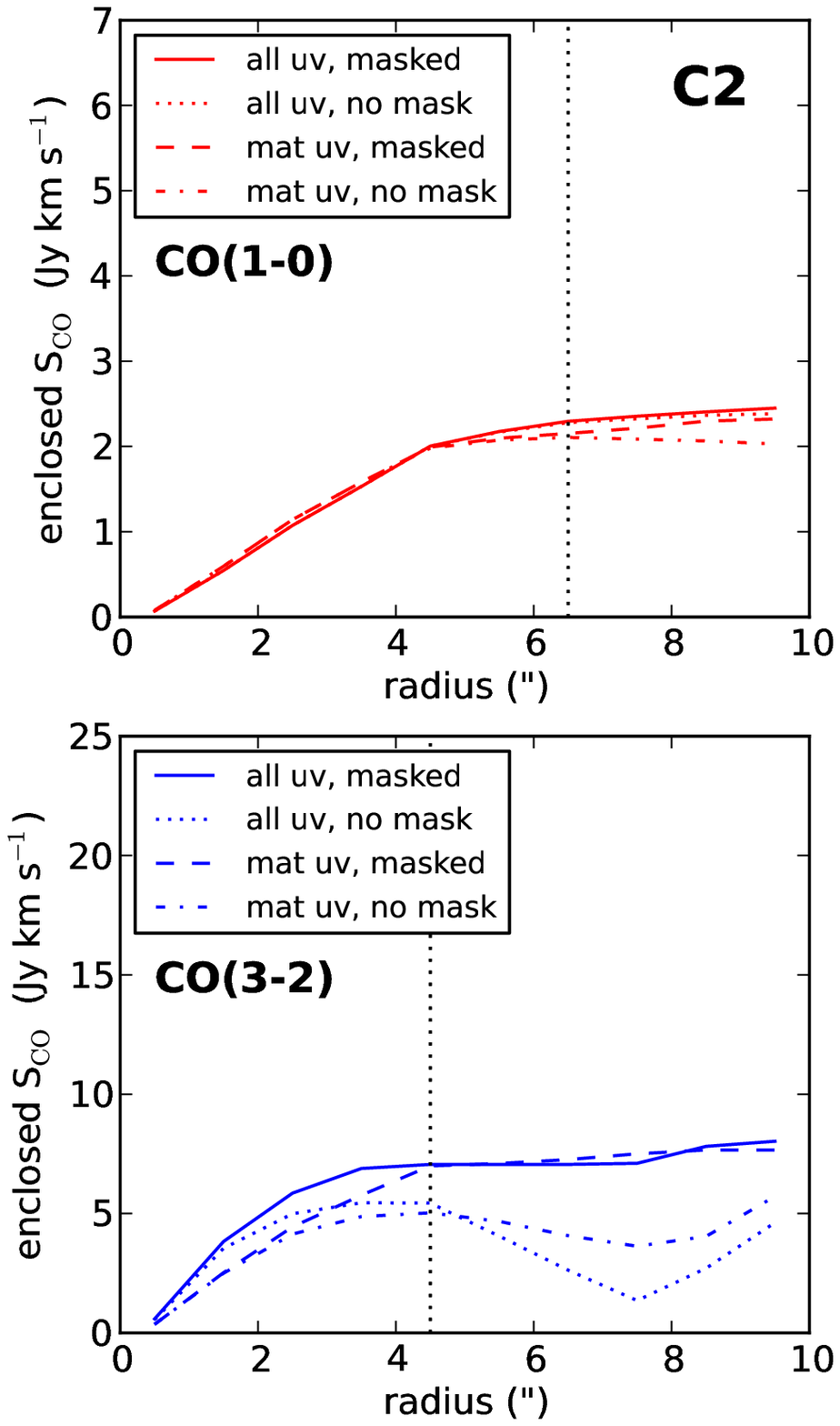}
\end{minipage}
\begin{minipage}[h]{0.32\linewidth}
\includegraphics[width=\linewidth]{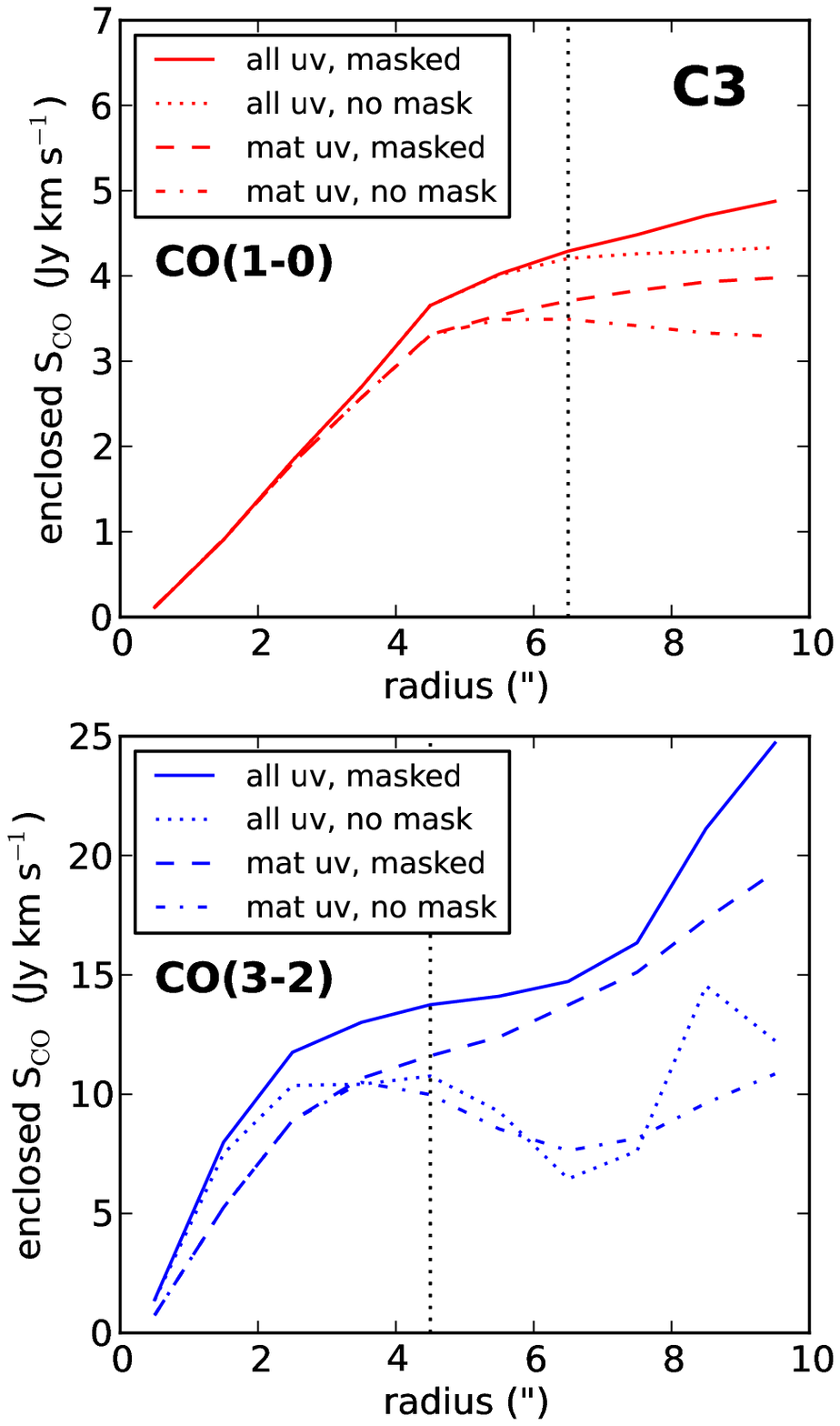}
\end{minipage}
\begin{minipage}[h]{0.32\linewidth}
\includegraphics[width=\linewidth]{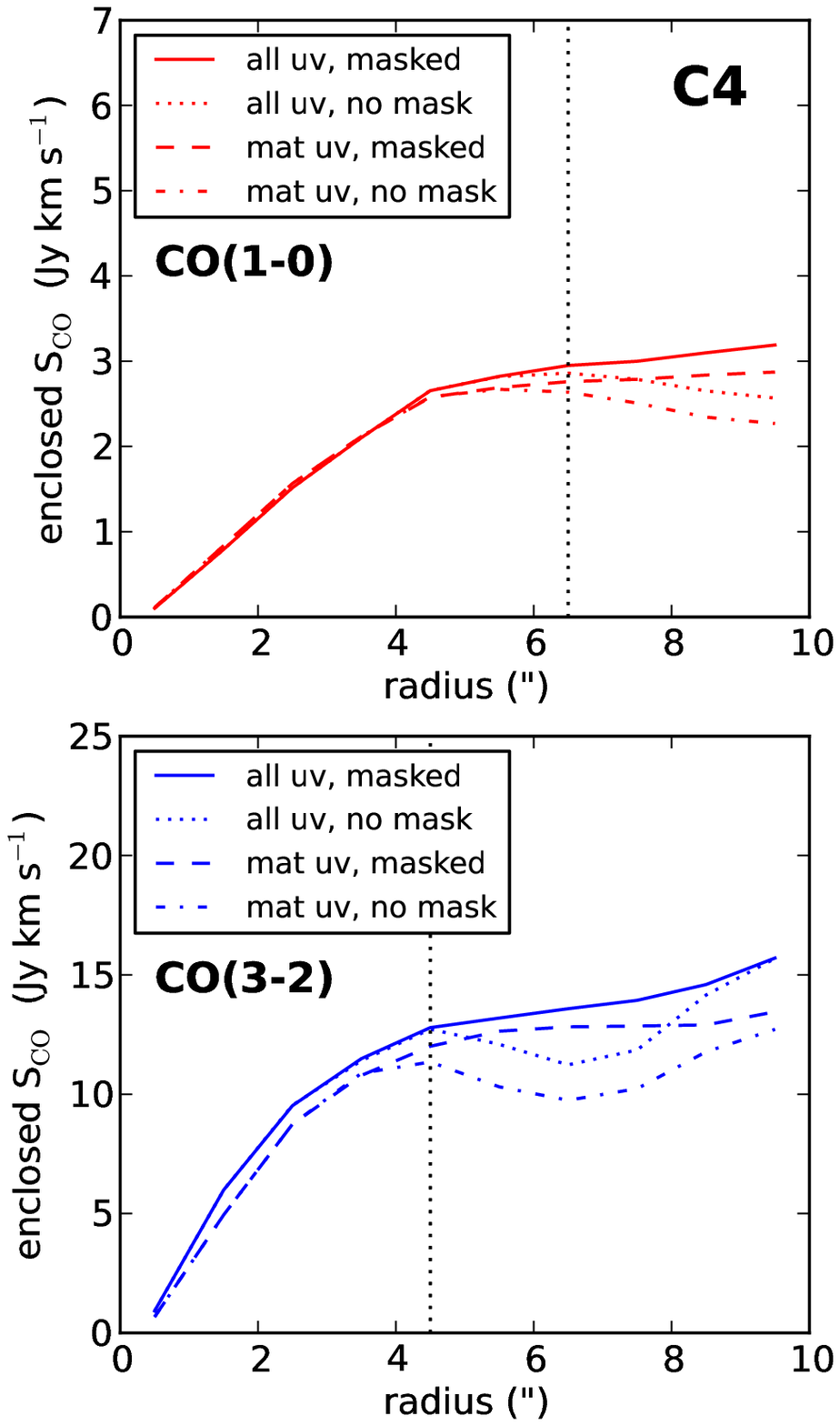}
\end{minipage}
\caption{Enclosed flux ($S_\mathrm{CO}$ in Jy \kms) as a function of radius for sources C2, C3 and C4.
The top panels show the CO\jone line flux for four cases: all $uv$ data, masked (solid) and unmasked (dotted); 
matched $uv$ data, masked (dashed) and unmasked (dash-dotted). The bottom panels show the CO\jthree line
flux for the same four cases. The vertical black dotted lines indicate the radius at which the enclosed flux peaks
($6.5\arcsec$ for CO\jone, $4.5\arcsec$ for CO\jthree). }
\label{fig:radprofile}
\end{figure*}

\subsection{Flux Estimation}
\label{sec:fluxest}
Total source fluxes in the CO lines are calculated by summing `source' pixels
in each `source' velocity plane of the image. The source velocity planes are 
selected by eye. The source pixels are those within the `source' region
which are not masked by our smooth mask (adapted from \citealt{Dame2011}). 
The smooth mask 
is created by applying a $2\sigma$ clip to a smoothed version of the image: 
Hanning smoothing is done along the velocity axis, and each velocity plane
is convolved with a Gaussian beam twice the size of the original synthesized beam. This smoothed
mask is used in order to exclude noise pixels but still capture low-level emission that a simple clip would miss.
In our own testing (using a model source of known flux inserted
into real data), we found a $2\sigma$ clip to
best reproduce the true flux over a range of detection significance levels. 

The appropriate source region size is selected to recover all of the flux without including the negative bowl. 
Since we do not have single dish data to complement the interferometric data presented here, 
our datasets are missing the shortest $uv$ spacings. As a result, the emission in the clean images
sits in a negative bowl, which will affect the measured flux (calculated by summing pixels within a given radius).
In order to accurately estimate the flux, 
we use the radius at which the radial profile of the enclosed flux first peaks, thereby excluding the negative bowl. 
The radial profiles of the enclosed flux for sources C2, C3 and C4 are
shown in Figure \ref{fig:radprofile}: CO\jone in red in the top panels and CO\jthree in blue in the bottom panels.
The profiles for both $uv$ selections (all and matched), unmasked and masked are shown for each transition.
The negative bowl is most evident in the unmasked profiles (dotted and dash-dotted lines), in which the enclosed flux 
peaks and then decreases with radius. The radius of the first peak of the enclosed flux distribution is
$6.5\arcsec$ for the CO\jone data and $4.5\arcsec$ for the CO\jthree data (shown by the vertical black dotted lines).
These radii are used in the calculation of total fluxes throughout this work.

The error in the flux measurement is estimated from the standard deviation of the measured fluxes
using different velocity channel averaging and starting channel, using three different methods 
of calculating the flux in each case.
The three methods are: the $2\sigma$ masking technique described above, 
the same masking technique with a $3\sigma$ clip, and the simple addition of all pixels (no mask) within the
source region. 

We performed extensive testing of our analysis technique in order to choose the parameters of the reduction
to eliminate systematic offsets and minimize the uncertainty 
due to noise (as described above). 
We find a 10-30\% error in the flux measurement coming from noise in the data reduction and analysis steps. 
From this, we take an average uncertainty of 20\% in the flux estimated in each channel, which results in a
uncertainty in the total flux of 20\%$(N_\mathrm{ch})^{-0.5}$ ($N_\mathrm{ch}$ is the number of velocity
channels in which the flux is summed). For the total flux values reported in Table \ref{tab:COprop}, we estimate
the error from the variation in the flux calculated using different channel averaging, flux measurement method, etc. 
(described above), which is consistent with the 20\%$(N_\mathrm{ch})^{-0.5}$ we expect. 
In our analysis of the integrated flux velocity profiles (Section \ref{sec:carefulr31}), we assume errors of 20\%
in the flux in each channel.

Further, these data suffer from systematic errors due to absolute flux calibration and primary beam correction.
We set the flux scale in our dataset based on a primary flux calibrator (Mars or MWC349), the flux of which
is only known to $\approx20\%$. In the primary beam correction of the dataset, pointing and focus errors at the 
time of the observations as well as errors in the primary beam model can significantly reduce image fidelity, leading
to errors in the measured fluxes of $\approx20\%$ (see SKA Memo \#103, \citealt{SKAmemo103}). 
Combining these systematic errors in 
quadrature, we estimate that our flux measurements suffer from systematic uncertainties of up to $\approx30\%$. 
We consider all these factors in the presentation of our data in Table \ref{tab:COprop}: 
for the line flux ($S_\mathrm{CO}$), we report the measured error; for $L_\mathrm{CO}'$, we include a $30\%$ systematic error
(added in quadrature to the measured error in $S_\mathrm{CO}$).

\newpage

\begin{figure*}[t]
\centering
\includegraphics[width=\linewidth]{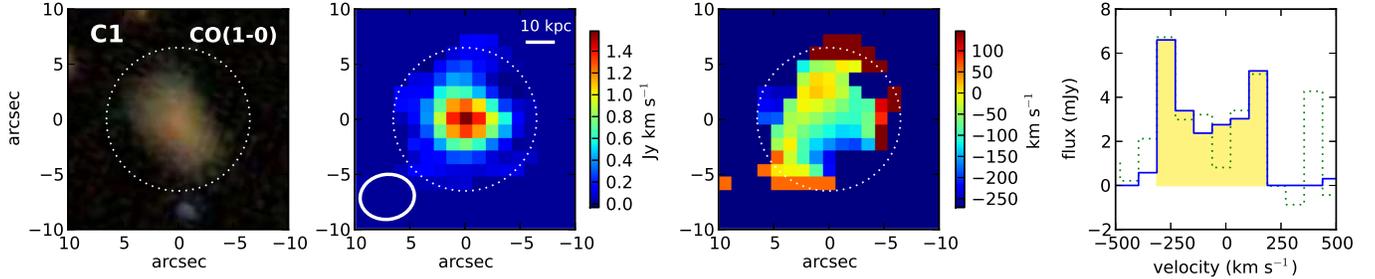}
\caption{CO\jone emission in source C1. The left panel shows the optical image.
The moment 0 and moment 1 maps are displayed in the left middle and right middle panels respectively. 
The dotted white ellipse indicates the source region (6.5\arcsec radius). 
In the moment 0 map, the beam size is indicated by the solid white ellipse in the lower left corner 
and a 10 kpc scale bar is given in the top right.
The far right panel shows the spectrum of the galaxy: the solid blue line is calculated with masking, 
the dotted green line is without.}
\label{C1fig}
\end{figure*}

\section{Moment Maps}
\label{sec:mommaps}
\begin{figure*}[p]
\centering
\includegraphics[height=6.2in,angle=-90]{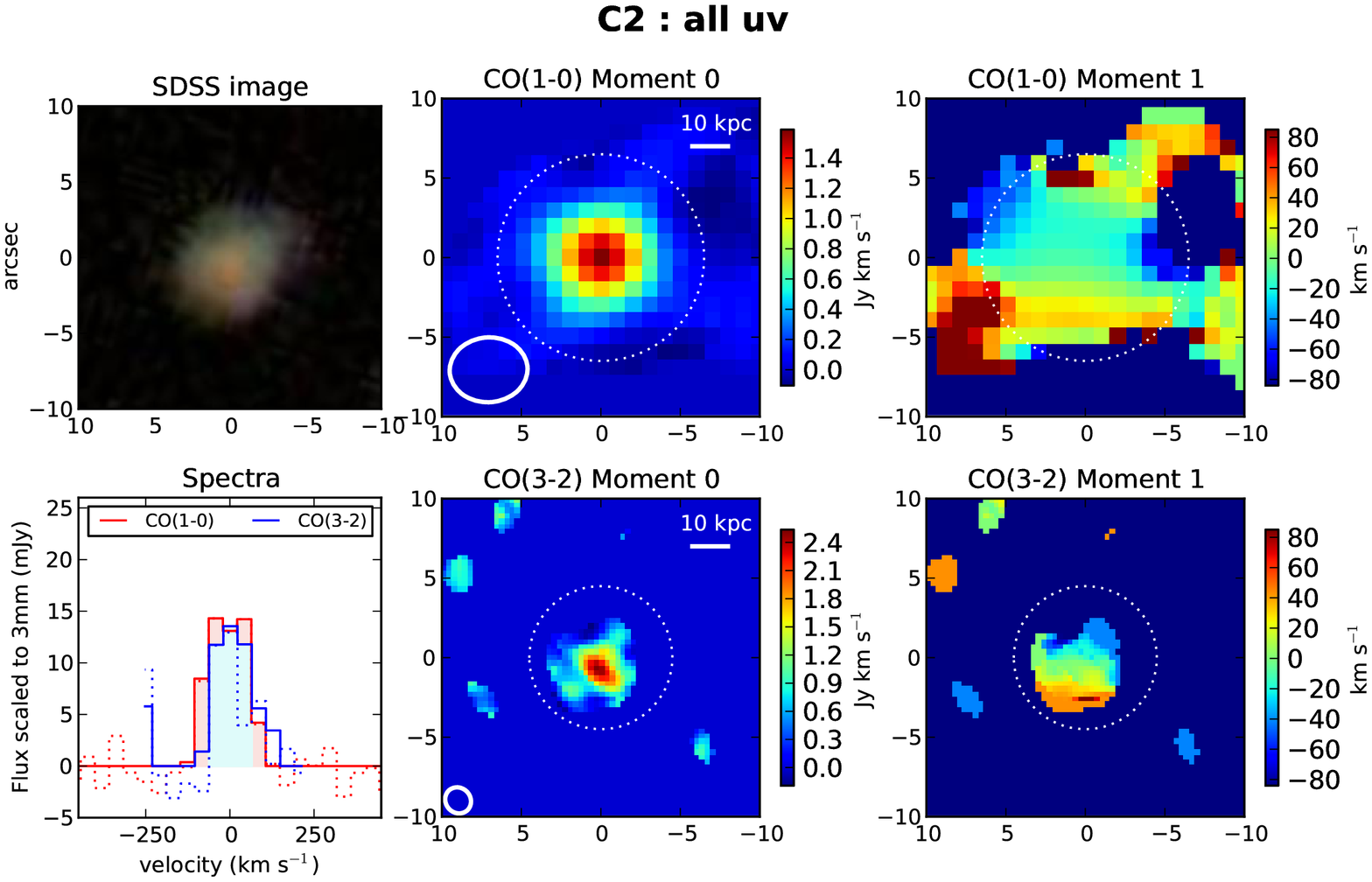}
\rule[-0.1cm]{6in}{0.01cm}
\includegraphics[height=6.2in,angle=-90]{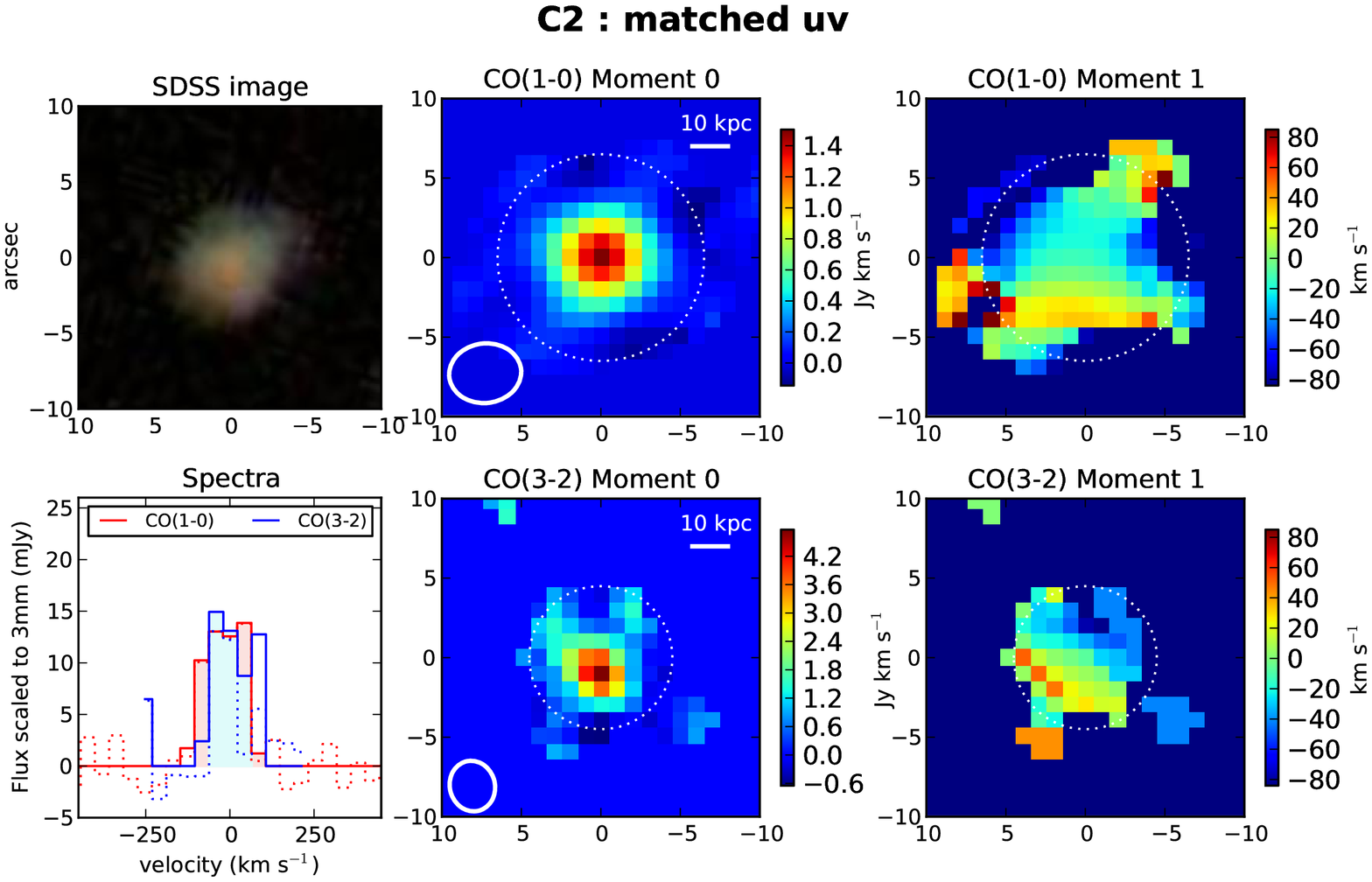}
\caption{Two sets of 6 panels showing the optical image (top left), moment maps (middle and right panels) 
and spectra (bottom left) 
for source C2: the top set uses all $uv$ data, the bottom set uses matched $uv$ data. In each set, the top
middle and right panels show the moment 0 (total intensity) and moment 1 (intensity-weighted mean velocity)
maps, respectively, for the CO\jone line. The bottom middle and right panels show the same, but for the CO\jthree line.
The synthesized beam (solid white ellipse) and 10 kpc scale (white bar) are indicated in each middle panel.
The dotted white circles in the middle and right panels show the source regions in which flux is summed 
(with radii of 6.5\arcsec and 4.5\arcsec for the CO\jone and CO\jthree data respectively; see Section \ref{sec:fluxest}).
The bottom left panels show the spectra in mJy, with the CO\jthree spectrum reduced by a factor of 4.5 to 
match the scale of the CO\jone spectrum.}
\label{C2fig}
\end{figure*}

\begin{figure*}[p]
\centering
\includegraphics[height=6.2in,angle=-90]{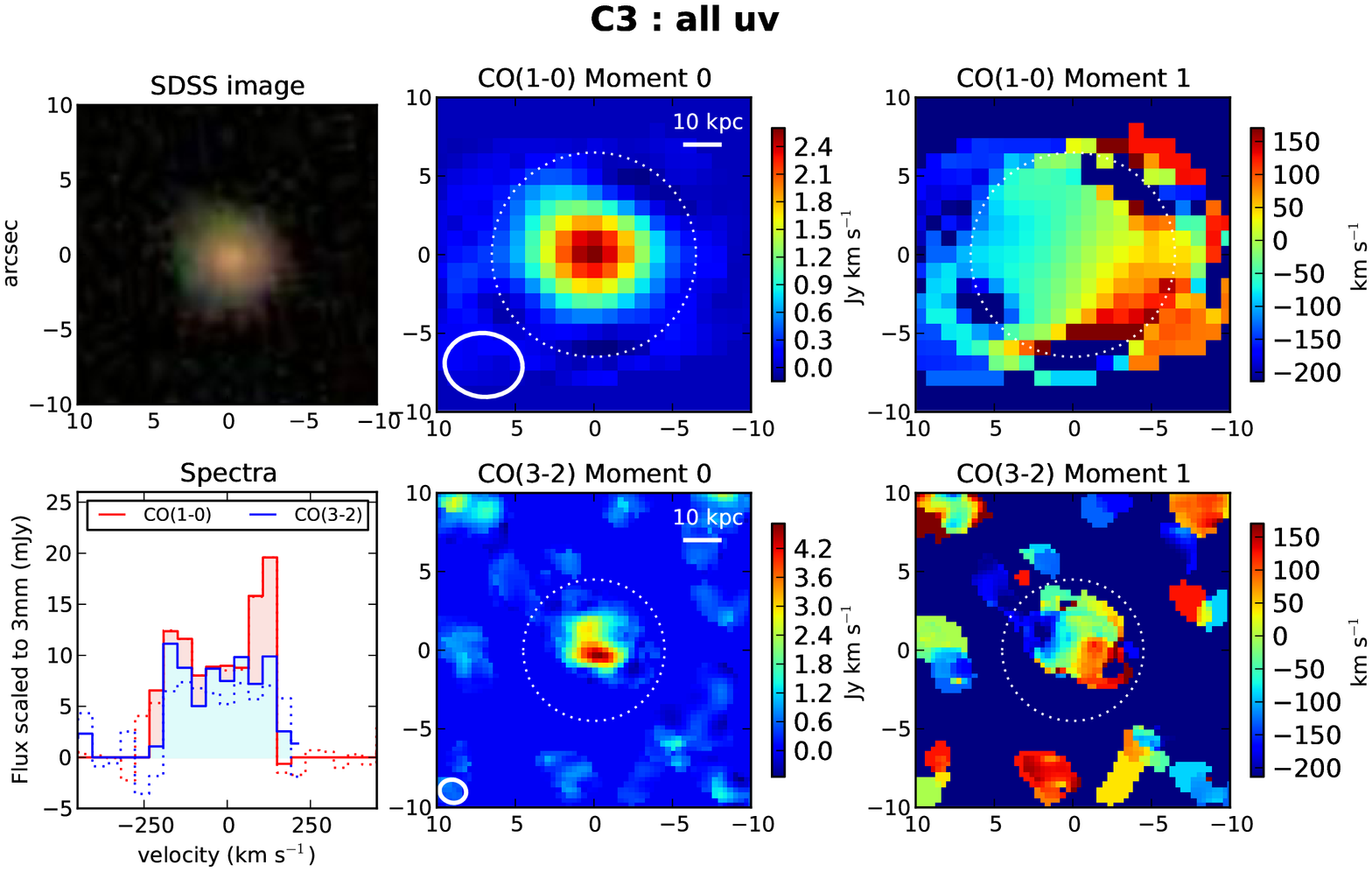}
\rule[-0.1cm]{6in}{0.01cm}
\includegraphics[height=6.2in,angle=-90]{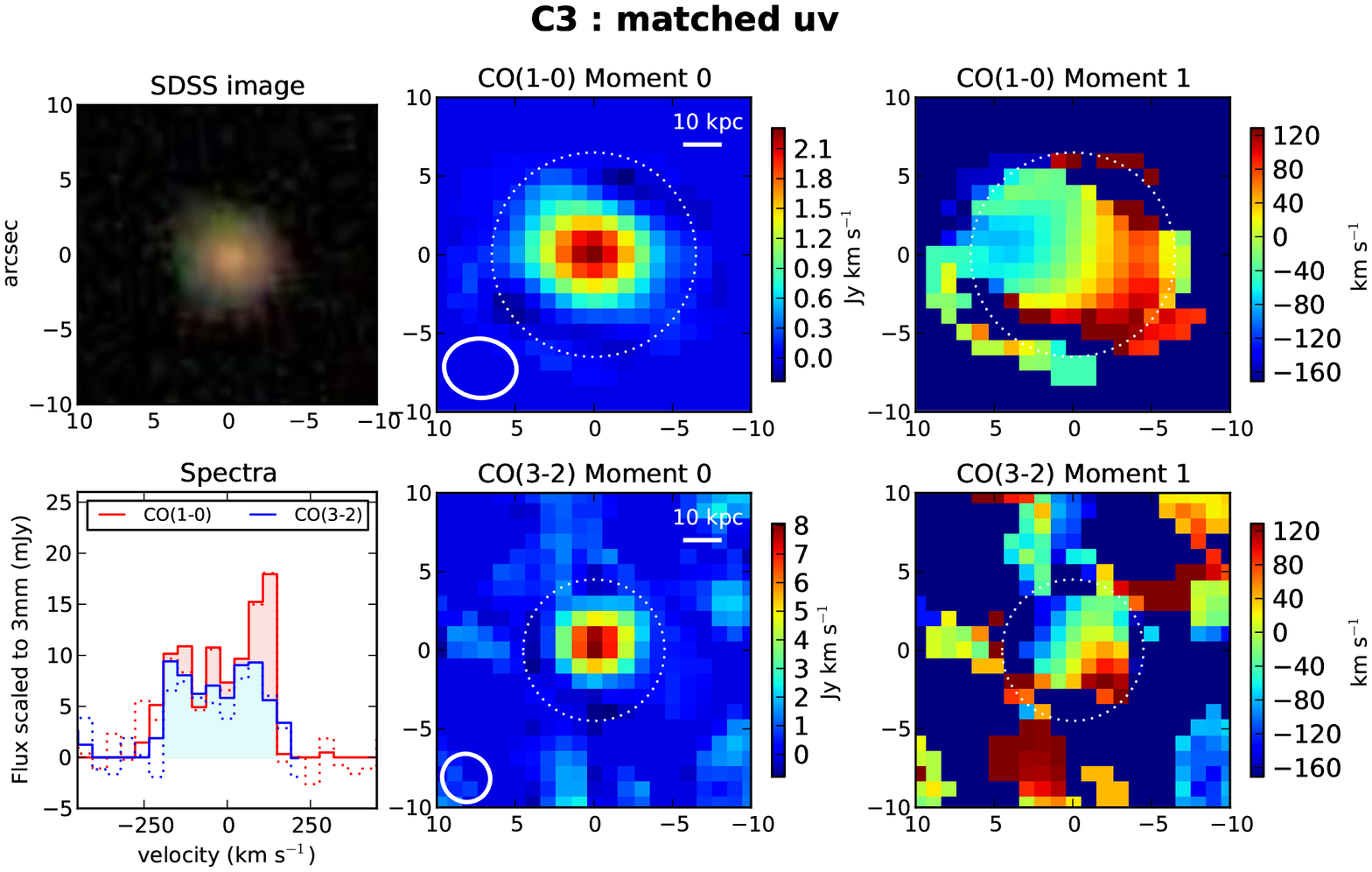}
\caption{Same as Figure \ref{C2fig}, but for source C3.}
\label{C3fig}
\end{figure*}

\begin{figure*}[p]
\centering
\includegraphics[height=6.2in,angle=-90]{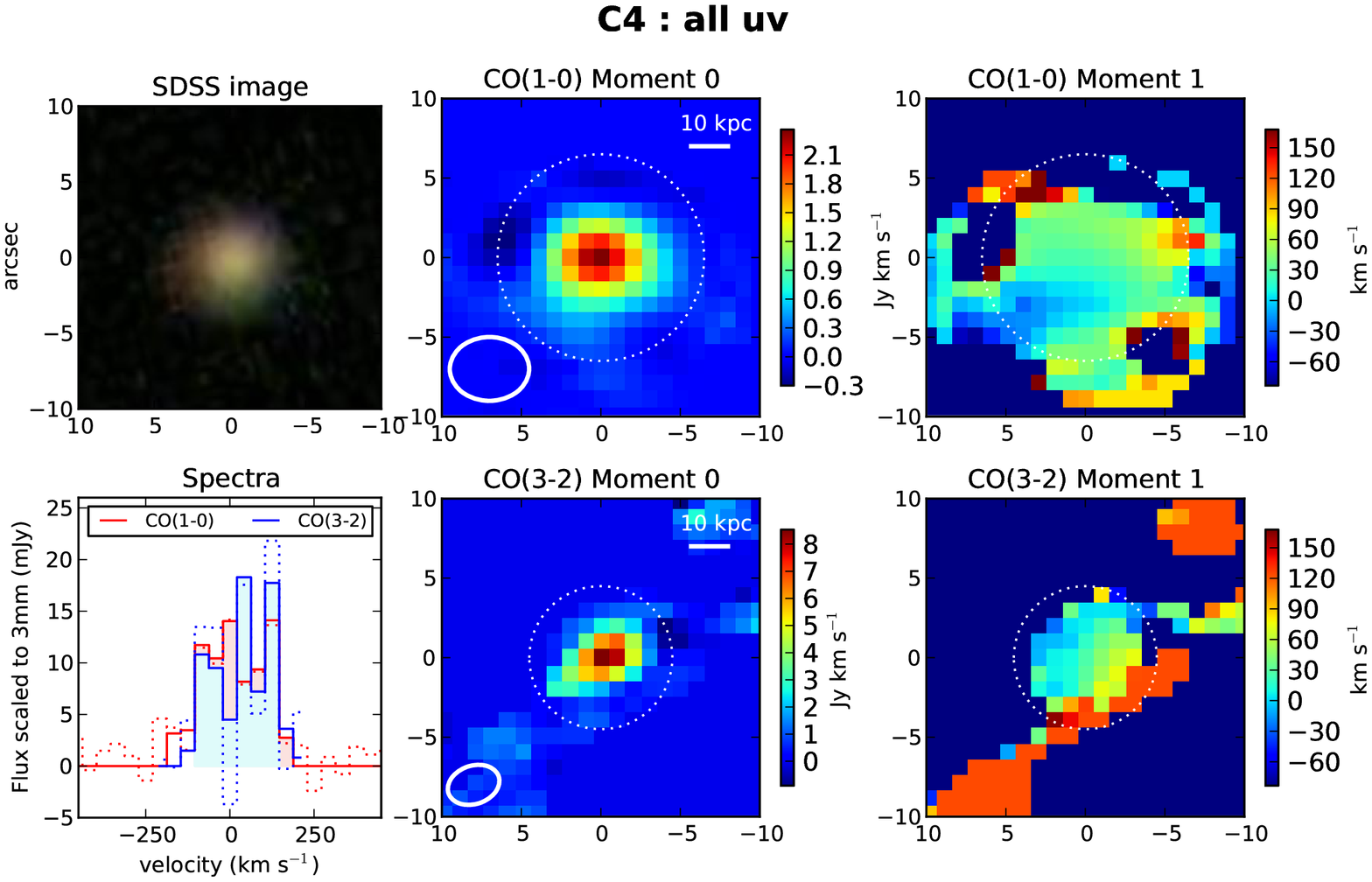}
\rule[-0.1cm]{6in}{0.01cm}
\includegraphics[height=6.2in,angle=-90]{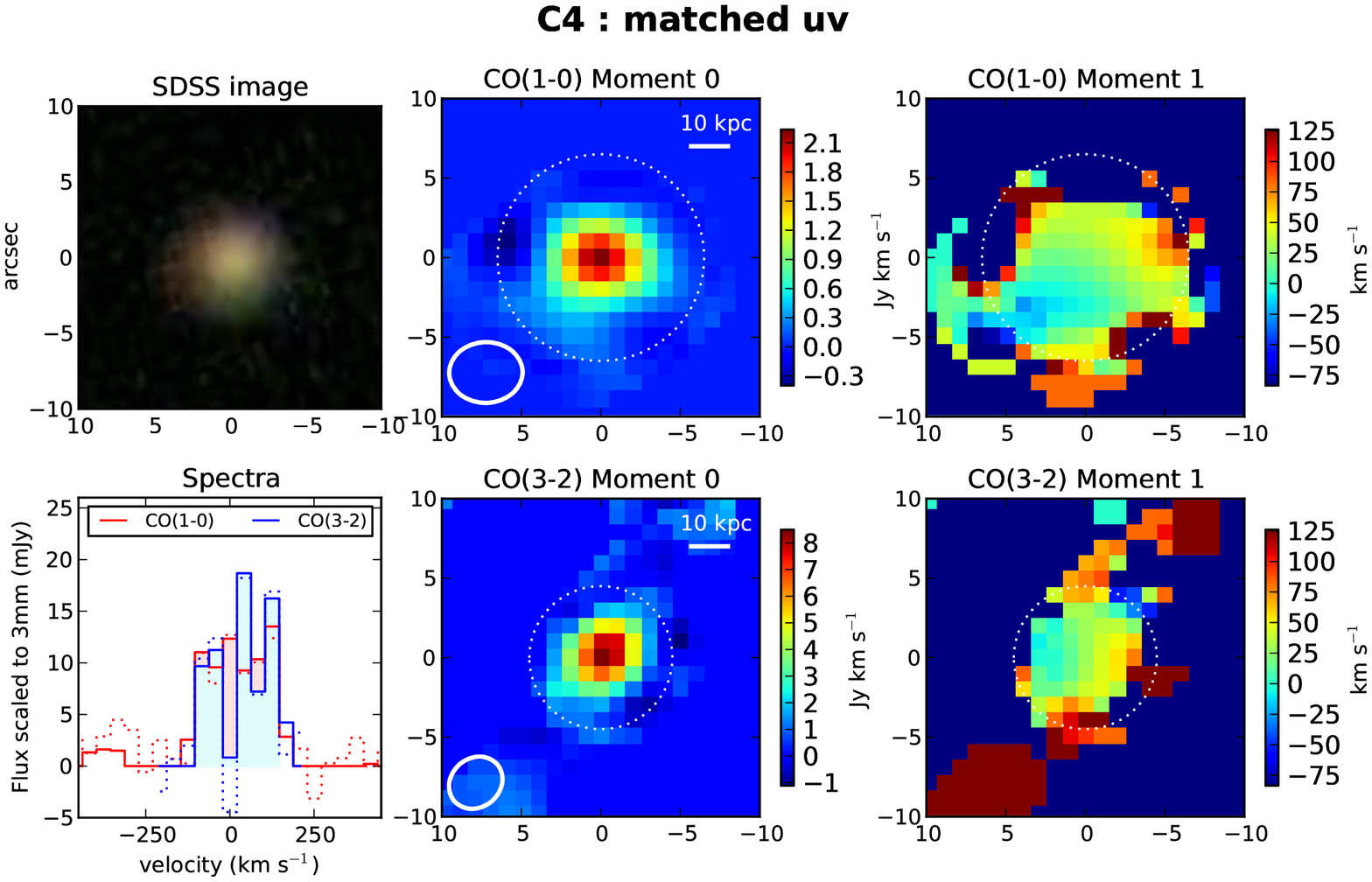}
\caption{Same as Figure \ref{C2fig}, but for source C4.}
\label{C4fig}
\end{figure*}

In this appendix, we present the moment maps, optical images and spectra of the CO emission
in the EGNoG bin C galaxies. Figure \ref{C1fig} shows the CO\jone emission in galaxy C1
(we do not detect CO\jthree emission in galaxy C1).
Figures \ref{C2fig} - \ref{C4fig} show both CO\jone and CO\jthree emission
in galaxies C2, C3 and C4, using all $uv$ data and matched $uv$ data.

\section{Radial Dependence of $r_{31}$}
\label{sec:raddepappendix}

Since $r_{31}$ traces the local excitation conditions of the molecular gas of a galaxy, it is expected
to vary within the disk of the galaxy. In fact, \citet{Dumke2001} found CO\jthree emission to be more
centrally concentrated than CO\jone emission in nearby galaxies, so that $r_{31}$ decreased with radius. 
To look for radial variation in $r_{31}$ in the EGNoG data, we cannot compare the emission maps directly 
due to the marginal resolution of the galaxies and the different $uv$ coverage of the two transitions. 
We emphasize that the $uv$ coverage (and thus different spatial resolution) of each transition
determines the shape of the radial profile of the enclosed flux (see Figure \ref{fig:radprofile} in Appendix \ref{sec:datreducandflux}), 
making a direct ratio of the two radial profiles meaningless without perfectly matched $uv$ sampling (even our 
matched $uv$ data do not meet this criterion due to different sampling within the allowed $uv$-distance range).

In order to disentangle the true emission distribution from the $uv$ sampling, we fit a two-dimensional Gaussian 
to the total intensity (moment 0) map of each CO transition using the MIRIAD program {\tt imfit}. 
The program fits for the position, peak intensity, total flux, size, and deconvolved source size. 
We perform the Gaussian fit on four versions of the moment 0 map for each transition in each galaxy:
for each $uv$ data selection (all and matched), we use moment 0 maps produced with and without 2$\sigma$ smooth masking.
%==============THESIS======================
%The fit position offsets, deconvolved size and total flux for each case are presented
%in Table \ref{tab:gaussianfit} along with the total flux calculated by a simple summing of pixels within the source region.
%The last column gives the percentage difference between the fit total flux and the summed total flux ($(\mathrm{fit} - 
%\mathrm{summed}) / \mathrm{summed})$). In two cases, {\tt imfit} found the deconvolved source to be consistent with 
%a point source: this is indicated by a `-' in Table \ref{tab:gaussianfit}. 
%==============THESIS======================
In order to derive a good fit to the source emission, we restrict the Gaussian fit to the standard source 
region (Section \ref{sec:fluxest}).
The error in the fit total flux is calculated by propagating the errors on the fit peak intensity and (not deconvolved) size.
Since these parameters are not independent, this error is likely an overestimate of the true error in the fit total flux.

The residual of each fit is inspected. We note that in general, the emission is fairly well fit by a Gaussian except for the CO\jthree
transition in galaxies C2 and C3, which are observed at higher resolution. The structure present it in these higher resolution images
is not well-fit by a Gaussian.
%==============THESIS======================
%, which is reflected in the large deviations of the fit total flux from the summed flux, especially
%when no masking is used.
%==============THESIS======================

We look for a radial dependence in $r_{31}$ by comparing the ratio calculated from the fit total fluxes to the ratio calculated
from the deconvolved peak intensities. The results are presented in Table \ref{tab:r31peak}. For each galaxy, for each combination
of $uv$ data selection (all, matched) and masking (no mask, masked), we derive the deconvolved peak brightness temperature
($T_\mathrm{b}$) for each CO transition and calculate $r_{31}$ from the deconvolved peak intensities as well as the fit total fluxes
(all values and errors  are derived from the parameters of the Gaussian fitting)
%==============THESIS======================
%, given in Table \ref{tab:gaussianfit}). 
%==============THESIS======================
The rest-frame peak brightness temperature (in K) is given by
\begin{equation}
T_\mathrm{b} = \frac{h \nu_0}{k_\mathrm{B} \ln\left[\frac{2 h \nu_0^3}{c^2 (1+z)^3 I_\mathrm{peak}} + 1\right]}
\end{equation}
where $\nu_0$ is the rest frequency of the transition and $I_\mathrm{peak}$ is the observed peak specific intensity (erg s$^{-1}$ cm$^{-2}$ Hz$^{-1}$ Sr$^{-1}$).
We calculate $I_\mathrm{peak}$ from the velocity width ($\Delta V$ in \kms), fit total flux ($S_\mathrm{CO}$ in Jy \kms) and fit deconvolved size 
(FWHM$_\mathrm{major}$ and FWHM$_\mathrm{minor}$ in arcseconds):
\begin{equation}
I_\mathrm{peak} = \frac{4 (\ln2) (206265^2) S_\mathrm{CO}}{\pi (10^{23}) \Delta V \mathrm{FWHM}_\mathrm{major} \mathrm{FWHM}_\mathrm{minor}} 
\end{equation}

The ratio of the lines, $r_{31}$, is calculated from the line flux (total or peak) according to Equations \ref{LCOdef} and \ref{r31equ}. 
Note that this ratio is not equivalent to the ratio of the brightness temperatures as defined here (see discussion in Section \ref{sec:discussion}).

The errors in the quantities given in Table \ref{tab:r31peak} are calculated 
from the errors in the relevant fit parameters reported by {\tt imfit}. 
%Since {\tt imfit} 
%does not report an error the deconvolved source sizes (FWHM$_\mathrm{major}$ and FWHM$_\mathrm{minor}$), 
We estimate the error in the deconvolved source sizes (FWHM$_\mathrm{major}$ and FWHM$_\mathrm{minor}$)
from the spread in the four estimates (all and matched $uv$, with and without masking) for each transition, for each galaxy.
For each quantity, we report the average value for each galaxy, taking the error to be the average fractional error. 
From the average values, we report the mean (with an uncertainty from the errors in the average values)
and standard deviation for the sample as a whole.

\FloatBarrier

We find peak brightness temperatures of approximately 2 K and 4 K for CO\jone and CO\jthree respectively. 
This is lower than the excitation temperature observed in Giant Molecular Clouds (GMCs) in the Milky Way 
\citep[e.g. 10-30 K,][]{Polychroni2012}, as expected since the filling factor for GMCs in the molecular gas disk is less than unity.

While the value of $r_{31}$ is a complex problem (discussed in detail in Section \ref{sec:discussion}), 
the larger brightness temperature of the CO\jthree line (relative to the CO\jone line) may be explained simply 
by differing excitation temperatures or filling factors.
For instance, if the CO\jone traces an additional diffuse, lower-excitation gas component not traced by the CO\jthree line (as observed
in SMGs by \citealt{Carilli2010, Riechers2011z3SMGs}), the average brightness temperature of the CO\jthree
line will be higher than the CO\jone line.
On the other hand, a larger brightness temperature in the CO\jthree line may be
due to a larger filling factor for the CO\jthree emission.
Our Gaussian fits find systematically smaller sizes for the CO\jthree emission relative to the CO\jone sizes, which
is consistent with the results of \citet{Dumke2001}, who found CO\jthree emission to be more centrally concentrated 
than CO\jone emission in nearby galaxies. Since the molecular gas disks of galaxies tend to have an exponential 
radial profile \citep[e.g.][]{Regan2001, Leroy2009}, the gas-rich central region of the galaxy will
account for a larger fraction of the CO\jthree emission area, increasing the effective filling factor of the CO\jthree 
emission relative to that of the CO\jone emission. Assuming a constant excitation temperature for both transitions, 
the larger effective filling factor of CO\jthree emission would produce a higher brightness temperature, as we observe.

While the error bars are large, we do find the ratio of the deconvolved peak intensities (representative of
the conditions in the central region of the galaxy) to be systematically higher than the ratio of the total
fluxes. In our sample, we find $r_{31}(\mathrm{peak}) = 0.73\pm0.31$, with a standard deviation across the three galaxies
of 0.04. We compare this to the ratio calculated from total fluxes: $0.42\pm0.11$, with a standard deviation of 0.04.
While the small standard deviations in our $r_{31}$ values show consistency between the three galaxies, 
these results are plagued by sizable uncertainties as a result of fitting data with only modest signal to noise
(note that our error estimates are conservative and may in fact be over-estimating the true error).
Therefore, we conclude (but not robustly) that $r_{31}$ is higher in the center of the EGNoG galaxies 
than in the molecular disk as a whole \citep[consistent with][]{Dumke2001}.

\begin{table}[h]
\centering
\renewcommand{\arraystretch}{1.2}
\begin{tabular}{| c | c | c | c | c | c |}
\hline
Na- &  $uv$ & \multicolumn{2}{ c|}{Gaussian Fit Peak $T_\mathrm{b}$} & \multicolumn{2}{c|}{Gaussian Fit $r_{31}$} \\
\cline{3-6}
me & Sel. & CO(1-0) & CO(3-2) &  Peak &  Total \\
\Cline{0.8pt}{1-6}
C2 & all & $ 1.83\pm0.32$ & $ 3.79\pm0.40$ & $ 0.75\pm0.54$ & $ 0.41\pm 0.26$ \\
   &        & $ 1.83\pm0.32$ & $ 3.71\pm0.34$ & $ 0.68\pm0.47$ & $ 0.48\pm 0.29$ \\
\cline{2-6}
   & mat & $ 1.91\pm0.13$ & - &  - & $ 0.37\pm 0.13$ \\
   &        & $ 1.90\pm0.15$ & $ 3.86\pm0.47$ & $ 0.72\pm0.42$ & $ 0.42\pm 0.20$ \\
\Cline{0.8pt}{2-6}
 & avg & $1.87\pm0.23$ & $3.79\pm0.40$ & $0.72\pm0.48$ & $0.42\pm0.22$ \\
\Cline{0.8pt}{1-6}
C3 & all & $ 1.78\pm0.32$ & $ 3.45\pm0.35$ & $ 0.53\pm0.40$ & $ 0.44\pm 0.14$ \\
   &        & $ 1.65\pm0.27$ & $ 3.51\pm0.51$ & $ 0.74\pm0.67$ & $ 0.38\pm 0.22$ \\
\cline{2-6}
   & mat & $ 1.68\pm0.28$ & $ 3.75\pm0.37$ & $ 0.94\pm0.67$ & $ 0.35\pm 0.07$ \\
   &        & $ 1.65\pm0.27$ & $ 3.65\pm0.38$ & $ 0.88\pm0.66$ & $ 0.36\pm 0.11$ \\
\Cline{0.8pt}{2-6}
 & avg & $1.69\pm0.28$ & $3.59\pm0.40$ & $0.77\pm0.60$ & $0.38\pm0.13$ \\
\Cline{0.8pt}{1-6}
C4 & all & $ 2.03\pm0.60$ & $ 4.10\pm0.40$ & $ 0.76\pm0.72$ & $ 0.48\pm 0.38$ \\
   &       & $ 2.09\pm0.33$ & $ 4.16\pm0.44$ & $ 0.74\pm0.46$ & $ 0.45\pm 0.16$ \\
\cline{2-5}
   & mat & $ 2.08\pm0.36$ & $ 3.89\pm0.31$ & $ 0.57\pm0.34$ & $ 0.50\pm 0.15$ \\
   &        & $ 2.23\pm0.37$ &  - &  - & $ 0.45\pm 0.19$ \\
\Cline{0.8pt}{2-6}
 & avg & $2.11\pm0.42$ & $4.05\pm0.38$ & $0.69\pm0.50$ & $0.47\pm0.22$ \\
\Cline{0.9pt}{1-6}
 \multicolumn{2}{|r|}{mean}  & $1.89\pm0.19$ & $3.81\pm0.23$ & $0.73\pm0.31$ & $0.42\pm0.11$ \\
 \multicolumn{2}{|r|}{$\sigma$}  & 0.21 & 0.23 & 0.04 & 0.04 \\
\hline
\end{tabular}
\caption{Gaussian fit $T_\mathrm{b}$ and $r_{31}$. 
Deconvolved, rest-frame peak brightness temperatures (peak $T_\mathrm{b}$) and $r_{31}$
calculated from the deconvolved peak intensities ($r_{31}$ peak) and from the total fluxes ($r_{31}$ total). 
All values are calculated using the parameters obtained from the Gaussian fitting.
For each galaxy, we report five values in each column: one for each of the four data selections used 
(all $uv$ (all) and matched $uv$ (mat), without masking first and with masking second) and the average 
value with a typical error. From the average values for each galaxy, we calculate
the mean (with an uncertainty from the errors in the average values) and the standard deviation for each
quantity, presented in the bottom rows.  }
\label{tab:r31peak}
\end{table}

\end{appendix}

\end{document}